\documentclass[fleqn,usenatbib]{mnras}

\usepackage{newtxtext,newtxmath}

\usepackage[T1]{fontenc}

\DeclareRobustCommand{\VAN}[3]{#2}
\let\VANthebibliography\thebibliography
\def\thebibliography{\DeclareRobustCommand{\VAN}[3]{##3}\VANthebibliography}

\usepackage{graphicx}	
\usepackage{amsmath}	
\usepackage{xcolor}

\title[The dark side of galaxy stellar populations]{ The dark side of galaxy stellar populations I: The stellar-to-halo mass relation and the velocity dispersion - halo mass relation}

\author[L. Scholz-D\'iaz et al.]{
Laura Scholz-D\'iaz,$^{1,2}$\thanks{E-mail: scholz@iac.es}
Ignacio Mart\'in-Navarro,$^{1,2}$
and Jes\'us Falc\'on-Barroso$^{1,2}$
\\
$^{1}$ Instituto de Astrofísica de Canarias, E-38205 La Laguna, Tenerife, Spain\\
$^{2}$ Universidad de La Laguna, Dept. Astrofísica, E-38206 La Laguna, Tenerife, Spain
}

\date{Accepted XXX. Received YYY; in original form ZZZ}

\pubyear{2021}

\begin{document}
\label{firstpage}
\pagerange{\pageref{firstpage}--\pageref{lastpage}}
\maketitle

\begin{abstract}
 The growth and properties of galaxies are thought to be closely connected to the ones of their host dark matter halos. Despite the importance of this so-called galaxy-halo connection, the potential role of dark matter halos in regulating observed galaxy properties remains yet to be fully understood. In this work, we derive the ages, metallicites and [Mg/Fe] abundances from optical spectra from the Sloan Digital Sky Survey of nearby central galaxies, and study them in terms of their host halos. We investigate how the scatter in the stellar-to-halo mass relation and the velocity dispersion - halo mass relation correlates with these stellar population parameters. In addition, we also study the differences when distinguishing between different galaxy morphologies and environments. We find that the ages and chemical enrichment of galaxies are not fully determined by their stellar masses or velocity dispersion, but also depend on the mass of the host halos. Our findings suggest that the velocity dispersion is the best proxy of the stellar population parameters with halo mass playing a secondary yet noticeable role. We interpret that the origin of the correlation between the scatter of these relations and the ages and metallicities might be related to different halo formation times.
\end{abstract}


\section{Introduction}
\label{intro}

In the  $\rm \Lambda$CDM paradigm galaxies form and evolve within dark matter halos \citep[e.g.,][]{white1978core}. These halos form as gravitationally collapsed regions of the density field, since their evolution is mainly driven by the dark matter gravitational field \citep{1976MNRAS.177..717W,1988ApJ...327..507F,1996Natur.380..603B}. 
In this sense, the assembly of dark matter halos is relatively well understood theoretically owing to large N-body simulations \citep[e.g.,][]{2005Natur.435..629S}. 
To first-order, galaxy formation is thought to be mainly driven by halo mass growth \citep[e.g.,][]{white1978core,1984Natur.311..517B}, yet galaxies are not regulated by dark matter alone, and the baryonic procceses involved (e.g. gas cooling and heating,  galaxy mergers, black hole growth, stellar and black hole feedback) also play a key role in their evolution. However, the physics involved in these processes is complex and highly non-linear, and the physical mechanisms that regulate the baryonic cycle of galaxies remain to be fully understood.

The physical and statistical connection between observed galaxies (i.e. the baryonic matter) with their host halos (i.e. the dark matter) is fundamental for understanding the physics of galaxy formation in this cosmological scenario \citep[for a recent review see][]{2018ARA&A..56..435W}. This  galaxy-halo connection has been long studied in the literature, yet how the growth and properties of galaxies (e.g. mass, morphology, color, size, star formation, gas fraction) depend on different halo properties (e.g. mass, formation time, concentration, environment, accretion rate, spin) \citep[e.g.,][]{2011MNRAS.410..210M,2013MNRAS.431..600W,2014MNRAS.444..729H,2017ApJ...834...37L,2017arXiv171110500H,2015MNRAS.446..651W,2018MNRAS.478.4487T,2019MNRAS.488.3143B} remains a matter of debate \citep[][]{2018ARA&A..56..435W}. 

One of the key aspects of this  connection is the stellar-to-halo mass relation (SHMR), which links the stellar mass of galaxies to the masses of their host halos. The mean SHMR has been derived using a variety of different methods and observables: direct and parameterized abundance matching \citep[e.g.,][]{2010ApJ...717..379B,2013ApJ...770...57B,2010ApJ...710..903M,2013MNRAS.428.3121M,2010MNRAS.404.1111G,2010MNRAS.402.1796W,2013ApJ...771...30R}, galaxy groups and clusters \citep[e.g.,][]{2004ApJ...617..879L,2009ApJ...695..900Y,2009ApJ...699.1333H,2018AstL...44....8K}, halo occupation distribution \citep[e.g.,][]{2007ApJ...667..760Z}, the conditional luminosity function \citep[e.g.,][]{2012ApJ...752...41Y}, and empirical modelling \citep[e.g.,][]{2019MNRAS.488.3143B}. The SHMR and its scatter have also been probed extensively using: satellite kinematics \citep[e.g.,][]{2007ApJ...654..153C,2011MNRAS.410..210M,2013MNRAS.428.2407W,2019MNRAS.487.3112L}, galaxy-galaxy weak lensing  \citep[e.g.,][]{2006MNRAS.368..715M,2016MNRAS.457.3200M,2015MNRAS.447..298H,2017MNRAS.467.3024L,2020MNRAS.499.2896T},  galaxy clustering \citep[e.g.,][]{2007ApJ...667..760Z,2013MNRAS.435.1313H,2016MNRAS.459.3040G,2019MNRAS.485.1196Z}, and a combination of these last two techniques \citep[e.g.,][]{2015MNRAS.454.1161Z,2016MNRAS.457.4360Z}. 

The scatter of the SHMR traces the efficiency of galaxy formation in dark matter halos \citep[see][]{2018ARA&A..56..435W}. It has been found that this scatter is connected to halo formation time or concentration using cosmological hydro-simulations and semi-analytical models \citep[e.g.,][]{2013MNRAS.431..600W,2017MNRAS.465.2381M,2017MNRAS.470.3720T,2018ApJ...853...84Z,2018MNRAS.480.3978A,2019MNRAS.490.5693B,2020MNRAS.491.5747M,2021arXiv210512145C}. These studies typically find that halos that formed earlier (or are more concentrated) host more massive galaxies than late-formed (or less concentrated) halos, i.e., they have been more efficient at forming stars. 
Observationally, \citet{2021MNRAS.505.5117Z} also found similar results using weak-lensing. 
The SHMR has also been studied in terms of galaxy colors or morphologies  \citep[e.g.,][]{2011MNRAS.410..210M,2013MNRAS.428.2407W,2015ApJ...799..130R,2016MNRAS.455..499L,2016MNRAS.457.3200M,2016MNRAS.457.4360Z,2019arXiv191104507T,2020MNRAS.499.4748M,2020MNRAS.499.3578C,2021arXiv210512145C}, yet how the scatter of the SHMR is related to these galaxy properties remains a matter of debate. At the moment, there is no agreement between different methods in whether red (or passive) galaxies are more-massive, less-massive or have the same mass than blue (or star-forming) galaxies, at a given halo mass \citep[for more details see section 6.1 of ][]{2018ARA&A..56..435W}.  

Stellar populations within galaxies retain information about their evolutionary paths, and thus, they are potentially linked to the properties and growth of the host dark matter halos. These observed galaxy properties, such as ages and metallicities, are fossil records of the past history of star formation and chemical enrichment of the galaxies. In the Local Universe these stellar population properties follow tight scaling relations with galaxy stellar mass, $M_{\star}$, or velocity dispersion, $\sigma$  \citep[e.g.,][]{1973ApJ...179..731F,1989PhDT.......149P,1992ApJ...398...69W,1993ApJ...411..153B,2000AJ....119.1645T,2005MNRAS.362...41G,2005ApJ...621..673T,2015MNRAS.448.3484M}, with galaxies becoming older and more-metal rich as $M_{\star}$ or $\sigma$ increases. In particular, for early-type galaxies it is also found that the element abundance ratio [$\alpha$/Fe] correlates with $M_{\star}$ and $\sigma$ 
\citep[e.g.,][]{1999MNRAS.306..607J,2000AJ....119.1645T,2000AJ....120..165T,2005ApJ...621..673T,2006MNRAS.370.1106G,2010MNRAS.408...97K,2015MNRAS.448.3484M}. The [$\alpha$/Fe] ratio can be thought as a proxy for the timescale in which the bulk of star formation occurs in a galaxy \citep{2005ApJ...621..673T,2011MNRAS.418L..74D}. If star formation in a galaxy is truncated early, it would end up with an $\alpha$-enhancement with respect to galaxies with more prolonged periods of star formation. 

In this sense, the present-day stellar content of these massive early-types indicates that they have assembled earlier and faster, and have reached a higher level of chemical enrichment than less massive galaxies. This apparent `downsizing' provides important constrains for galaxy formation models, which need physical mechanisms able to shut down star formation in these massive galaxies early on and keep them quiescent over cosmic time. The quenching of massive galaxies is still under debate, although the standard quenching mechanism implemented in state-of-the art cosmological simulations is black hole feedback \citep[e.g.,][]{2005Natur.433..604D,2005MNRAS.361..776S,2006ApJS..163....1H,2006MNRAS.370..645B,2007MNRAS.380..877S,2008MNRAS.391..481S,2015Natur.521..192P}. These quenching mechanisms are expected to play a key role regulating the star formation efficiency in massive halos, and hence, to be connected to the scatter of the SHMR.

In this work we aim to study the stellar population properties of nearby galaxies in a cosmological context by probing the scatter of these observables across the SHMR. As a novelty, we also introduce the velocity dispersion - halo mass relation (VDHMR) and study its scatter analogously as for the SHMR. Galaxy stellar population properties have been studied in terms of their host halos in order to assess how they are affected by their environment, distinguishing between central and satellite galaxies \citep[e.g.,][]{2010MNRAS.407..937P,2014MNRAS.445.1977L,2021MNRAS.502.4457G,2021MNRAS.500.4469T}. However, whether halo mass plays a role in regulating the stellar content of galaxies is still an open question. 

The paper is organized as follows: section \ref{sec:data} introduces the SDSS data and catalogues that we used, and our sample of galaxies. In section \ref{analysis} we describe the stellar population model ingredients, and the kinematic and stellar population analyses. The stellar population properties across the SHMR are shown in section \ref{sec:smhm}, and across the VDHMR in section \ref{sec:sigmh}.  We quantify the dependence of the stellar population properties on different parameters in section \ref{sec:par}. The dependence of the VDHMR on morphology and environment are presented in sections \ref{sec:morph} and \ref{sec:env}, respectively.  We discuss our results in section \ref{sec:discussion} and they are summarized in section \ref{sec:concl}.

\section{Sample and data}
\label{sec:data}

In order to study the stellar population properties of large sample of galaxies, we use spectroscopic data from the Sloan Digital Sky Survey Data Release 9 \citep[SDSS DR9;][]{2000AJ....120.1579Y,2012ApJS..203...21A}. The spectra are taken with 3$''$-diameter fibers pointed to the centers of the galaxies, have a wavelength coverage in the optical/NIR (3800-9200\AA) and a spectral resolution R$\sim$2000. 

With the aim of investigating galaxies in a cosmological context, in addition to spectra we also need dark matter halo mass estimates. For that, we cross-match the SDSS sample with the galaxy group catalog from \citet{2007ApJ...671..153Y} to obtain dark matter halo mass estimates. The catalog comprises galaxies with redshifts in the range 0.01 $\leq z \leq$ 0.20 with a redshift completness $C>0.7$. Briefly, group membership of the galaxies are determined through an iterative process using a halo-based group finder algorithm. Halo masses are estimated through abundance matching by ranking the total characteristic luminosity of the group, $L_c$, or total characteristic stellar mass, $M_{ c}$. In this sense, two different halo mass estimates are associated to the groups, i.e., halo mass estimates based on the ranking of $L_c$ or $M_{ c}$ \citep[for more details see][]{2007ApJ...671..153Y}. 

In order to connect galaxy properties to their host dark matter halos, we assign these halo masses to their central galaxies, as \citet{2007ApJ...671..153Y} provides a single halo mass estimate for each group/cluster. Thus, we focus our analysis in central galaxies only, defining the brightest galaxy in a galaxy group/cluster as the central one. Note that we select only galaxies with SDSS spectroscopic redshifts and both halo mass estimates available. For the main analysis we use the halo mass estimate based on $L_c$ instead of the one based on $M_{ c}$. We justify and discuss this choice in appendix \ref{sec:othercat}. We select centrals in clusters with more than 3 members in order to have more reliable halo mass estimates. 

In order to calculate the stellar masses through the stellar population analysis (see details in section \ref{an_sp}), we first need the galaxy absolute luminosities. For that, we cross-match our sample with the New York University Value-Added Galaxy Catalog  \citep[NYU VAGC;][]{2005AJ....129.2562B} and convert the K-corrected absolute magnitudes $M_r$ into luminosities $L_r$. \\

After performing the kinematic and stellar populations analyses (see section \ref{analysis}) we also apply cuts in order to avoid noisy data that might lead to unrealistic results, which are described in section \ref{analysis}. Our final sample consists of 8,801 nearby central galaxies with redshifts $0.01\!<\!z\!<\!0.2$, with stellar masses  $10^{9.7}\!<\!M_{\star} \rm [ M_{\odot}]\!<\!10^{11.8}$ and halo masses $10^{11.7}\!<\!M_{h} \rm [M_{\odot}]\!<\!10^{15.4}$.

In section \ref{sec:morph} we divide the sample into late- and early-type galaxies. For that, we use the morphological catalog presented in  \cite{2018MNRAS.476.3661D}, who classified about 670,000 SDSS galaxies with Deep Learning assigning a T-Type to each one of them, a number which is related to their morphological type. They denote galaxies with T-Type $\leq0$ as {\it early-types}, while refering to those with T-Type $>0$ as {\it late-types}. We crossed-match our final sample with this catalog and found 5,248 galaxies in common (819 late-types and 4,429 early-types).

\section{Analysis}
\label{analysis}

To extract the galaxy kinematics and stellar population properties we use the Penalized Pixel-Fitting algorithm \citep[pPXF, ][]{2004PASP..116..138C,2017MNRAS.466..798C} fed with the $\alpha$-enhanced MILES stellar population synthesis models (described in section \ref{an_miles}). In short, pPXF finds the linear combination of single stellar population (SSP) models that best reproduce the galaxy spectrum by convolving the SSP models with the line-of-sight velocity distribution (LOSVD) of the galaxy, which is parametrized by a Gauss-Hermite function \citep{1993ApJ...407..525V}. We first use pPXF to derive the stellar and gas kinematics, and then to perform the stellar population analysis. In the latter we kept the stellar kinematics fixed, as this reduces the velocity dispersion-metallicity degeneracy \citep{2011MNRAS.415..709S}. Both analyses are described below in sections \ref{an_kine} and \ref{an_sp}.

\subsection{MILES models}
\label{an_miles}
The SDSS galaxy spectra are compared to the $\alpha$-enhanced MILES stellar population synthesis models \citep{2010MNRAS.404.1639V,2015MNRAS.449.1177V} in our kinematic and stellar population analyses (described in sections \ref{an_kine} and \ref{an_sp}.). The models are based on the empirical  stellar library MILES \citep{2006MNRAS.371..703S, 2007MNRAS.374..664C,2011A&A...532A..95F}, combined with  corrections from theoretical stellar spectra  \citep{2005A&A...443..735C,2007MNRAS.382..498C}. They also employ solar-scaled and $\alpha$-enhanced BaSTI isochrones \citep{2004ApJ...612..168P,2006ApJ...642..797P}. These SSP models have ages ranging from 0.03 to 14 Gyr, metallicities [M/H] from -2.27 to +0.40 dex, and $\mathrm{[\alpha/Fe]}$ abundances from +0.0 to +0.4 dex. For our analyses we adopt a universal Kroupa IMF \citep{2001MNRAS.322..231K}. We selected models with 15 different ages sampled roughly logarithmically at younger ages, with a time interval between them of 1 Gyr at older ages (from 0.03 to 13.5 Gyr). They span 11 metallicities and have the two available $\mathrm{[\alpha / Fe]}$ abundances.

\subsection{Kinematic analysis}
\label{an_kine}
To derive the radial velocity ($V$) and velocity dispersion ($\sigma$), we use pPXF in the spectral rest-frame range from 4800 to 5400 \AA. Without using high-order Gauss-Hermite moments and including additive polynomials. These polynomials are used to account for template mismatch, as they change the individual line strengths \citep{2017MNRAS.466..798C}. For that, they can be included in this kinematic analysis, but they cannot be used to infer stellar population properties, as they could lead to a non-physical combination of SSPs. As this spectral range includes several nebular emission lines (e.g. $H_{\beta}$, [$O_{III}$] $\mathrm{\lambda\lambda}$4959, 5007, [$N_I$] $\mathrm{\lambda\lambda}$5198, 5200), we also extract the gas kinematics in the fitting process. We consider a single gas kinematic component (i.e, the gas emission lines are modelled with a single gaussian), except for the the Balmer emission lines. For these lines we consider a second additional component, i.e., they are modelled as the sum of two gaussians. As $H_{\beta}$ is very sensitive to age, this modelling provides a better fit for galaxies with stronger emission. 

As the galaxy extent covered by the aperture of the SDSS fiber depends on its redshift, we calculate the stellar velocity dispersion within 1 effective radius, $\sigma_e$. For that, we apply an aperture correction to the value obtained with pPXF, $\sigma$, following $\sigma_R / \sigma_e = (R/R_e)^{\gamma}$ \citep{2017A&A...597A..48F}, which describes the average $\sigma$ gradient of the galaxies. 

We use the SDSS photometric pipeline \citep{2002AJ....123..485S} to obtain the effective radius, which calculated in each band $(g,r,i)$ both fitting an exponential disk and a De Vaucoulers profile. For each galaxy, we select the highest likelihood profile as the best-fitting one, and compute $R_e$ as the average of the corresponding effective radii in the 3 bands. Galaxies without these effective radius measurements available are excluded from the analysis. We also calculate this correction using galaxy stellar masses \citep[eq. 33 from ][]{2003MNRAS.343..978S} to estimate the effective radius, and found consistent results. 

Finally, we use pPXF results to exclude galaxies whose spectrum does not allow a detailed stellar population analysis. In particular, we remove from our analysis galaxies that have velocity dispersion uncertainties (obtained with pPXF) higher than 10 km/s, or a signal-to-noise per pixel (S/N) lower than $\sim$10 in order to exclude outliers from the analysis and perform a reliable stellar population analysis. We checked that a higher S/N cut leads to similar results to the ones shown in the following sections. 

\subsection{Stellar populations analysis}
\label{an_sp}

Each SSP model has a single age, [M/H], $\mathrm{[\alpha/Fe]}$, and therefore using the weights of the pPXF best-fitting solution we are able to derive the mass-weighted average age, [M/H] and $\mathrm{[\alpha/Fe]}$ for each galaxy. We fit the spectra in the same spectral range as for the kinematic analysis, from 4800 to 5400 \AA, where absorption features senstitive to age and metallicity are concentrated. We note that the sensibility to $\alpha$-elements in this range is mainly dominated by the magnesium triplet 5176.7$\,$\AA. Hence, hereafter in the text we will use the notation [Mg/Fe] instead of $\mathrm{[\alpha/Fe]}$. In this case, we include multiplicative polynomials to account for dust reddening and calibration inaccuracies of the spectra. In the fitting process we kept the stellar kinematics fixed to the values obtained in our kinematic analysis (see section \ref{an_kine}). In this kinematic analysis the emission amplitude is coupled with the strength of $H_{\beta}$ absorption, and thus with the inferred age, as we included additive polynomials. For that reason, we fit again the nebular emission lines. We impose linear regularization  \citep{10.5555/1403886} to the pPXF solutions, as the recovery of the star formation history (SFH) from a galaxy spectrum is an ill-defined problem \citep{2006MNRAS.365...46O,2006MNRAS.365...74O,2017MNRAS.466..798C}. This means that fluctuations due to noise in the galaxy spectrum can notably change the solution. Regularization allows solutions with smooth variations between the weights of models with adjacent ages, metallicities and $\mathrm{[Mg /Fe]}$ values. To set the regularization parameter, we first select a galaxy with a typical signal-to-noise for our sample, and find the solution that corresponds to the smoothest one between the many degenerate solutions compatible with the data \citep[for more details see ][]{2017MNRAS.466..798C}. We kept this regularization parameter fixed for the whole sample. We note that the average stellar population parameters do not change much due to regularization. We find that there is a population of galaxies that are hitting the boundaries of the models (around +0.0 and +0.4 dex for the [Mg/Fe]), hence we remove those galaxies from the analysis. 

 In addition, mass-to-light ratios, (M/L), are computed for each model and are available in different photometric filters \citep{2010MNRAS.404.1639V}. For each galaxy, we use the best-fitting model to estimate a mass-weighted average $\rm (M/L)_r$. These are combined with k-corrected luminosities $L_r$ to obtain galaxy stellar mass estimates which are less prone to degeneracies than those based on galaxy colors. 

As a model-independent alternative to the stellar population properties obtained with pPXF, we also measure the strengths of different absorption features sensitive to age and metallicity (see appendix \ref{ap:indices}). In particular, we measure the age-sensitive index $\mathrm{H\beta_o}$ \citep{2009MNRAS.392..691C}, $\mathrm{Mgb5170}$ \citep{1994ApJS...94..687W}, which is an indicator of the Mg abundance, and we compute the iron-sensitive index $\rm <\!\!Fe\!\!>$ \citep{2000AJ....119.1645T} and the combined $\mathrm{[MgFe]^{\prime}}$ line-strength, which is sensitive to the total metallicity and relatively independent of [Mg/Fe] abundance ratio \citep{2003MNRAS.339..897T}.

\section{Stellar-to-halo mass relation}
\label{sec:smhm}

\begin{figure}
    \centering
    \includegraphics[scale = 0.42]{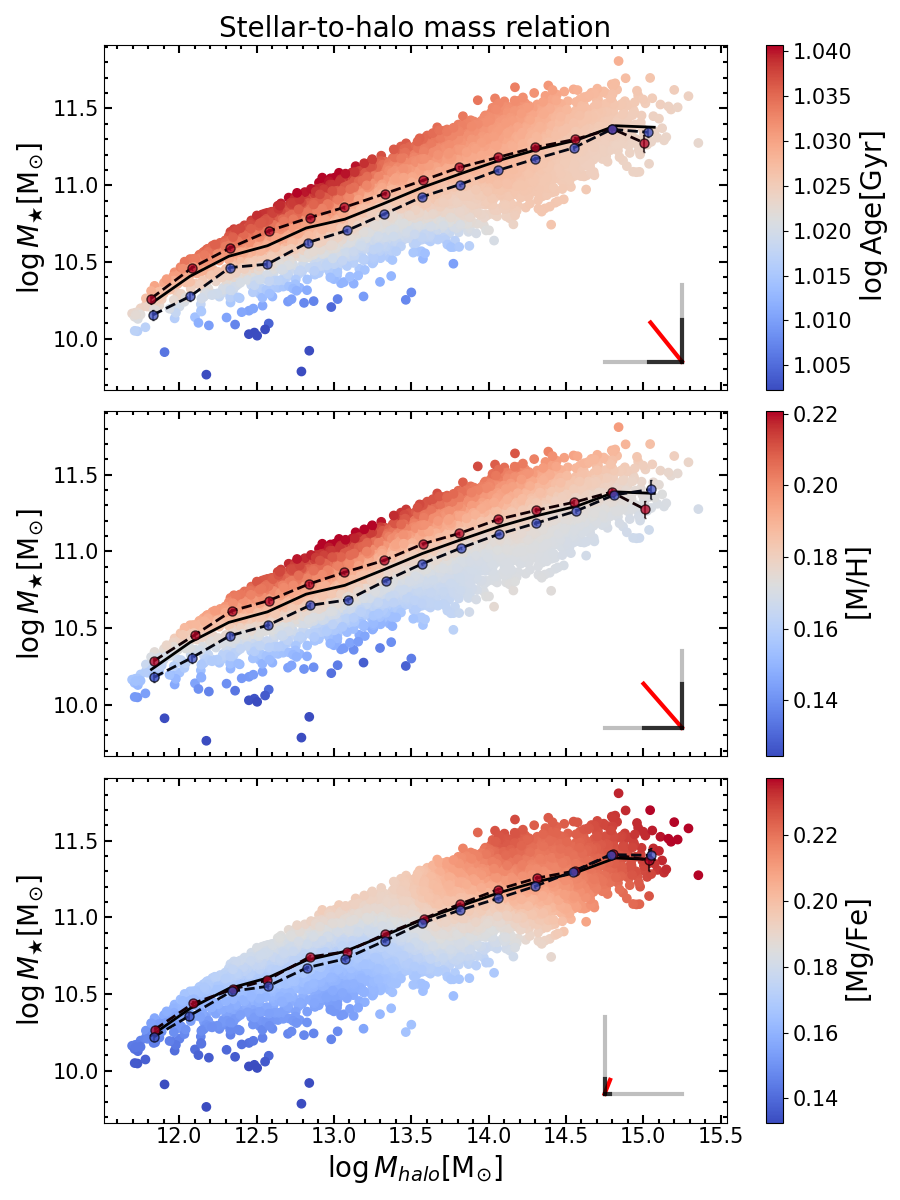}
    \caption{Stellar-to-halo mass relation for SDSS central galaxies color-coded by the stellar population properties resulting from our stellar population analysis described in section \ref{an_sp}: mass-weighted ages (upper panel), metallicities (middle panel) and [Mg/Fe] abundance (bottom panel). 
    We applied the LOESS algorithm to color-coding in order to highlight the global trends across the SHMR. The median SHMR is shown as a black solid line for reference. At fixed $M_h$, red and blue circles connected by dashed lines indicate the mean $M_{\star}$ of galaxies above the $\rm75^{th}$-percentile and below $\rm25^{th}$-percentile of the corresponding distributions. In the bottom right corner of the panels we show the partial correlation coefficient strengths (solid black lines) between the stellar population parameters and $M_{\star}$ (vertical) and $M_h$ (horizontal). For reference, the grey solid lines have a length which corresponds to a (Spearman) correlation coeffcient of 0.5. The red solid line indicates the direction of maximal increase of the stellar population parameters (see section \ref{sec:par} for more details).
    For low halo masses, the scatter correlates with the age and [M/H] at fixed $M_h$, while at fixed $M_{\star}$, galaxies in less massive halos are older and more metal-rich than the ones in more massive halos. At the high halo mass end, galaxies are mainly old and metal-rich.}
    \label{fig:smhm_pops}
\end{figure}

We first analyze the dependence of the stellar population properties of galaxies across the SHMR. In Fig. \ref{fig:smhm_pops}, we show the SHMR color-coded by the mass-weighted ages (upper panel), metallicities (middle panel) and [Mg/Fe] abundance ratios (bottom panel) resulting from our stellar population analysis described in section \ref{an_sp}. To highlight the global trends of the stellar population parameters across the SHMR, we applied the locally weighted regression (LOESS) algorithm of \citet{doi:10.1080/01621459.1988.10478639} to color-coding of Fig. \ref{fig:smhm_pops}\footnote{We use the LOESS implementation by  \citet{2013MNRAS.432.1862C}}. The median SHMR is shown as a black solid line for reference. At fixed $M_h$, red and blue circles connected by dashed lines indicate the mean $M_{\star}$ of galaxies above the $\rm75^{th}$-percentile and below $\rm25^{th}$-percentile of the corresponding distributions.

\subsection{Age and [M/H]} The ages and the metallicities of the galaxies show very similar trends over the SHMR, which interestingly seem to depend on the halo mass range considered. At low halo masses, below  $M_h$\,$\sim$\,10$^{13.5}$\,$\rm M_{\odot}$, the scatter of the SHMR shows a clear transition between younger and metal-poor for less massive galaxies to older and more metal-rich for more massive ones. 
At $M_h$\,$\sim$\,10$^{12.5}$\,$\rm M_{\odot}$, the median age and [M/H] of the galaxies with $M_{\star}$ values above the $\rm 75^{th}$-percentile of $M_{\star}$ is  $\sim$11 Gyr and $\sim$0.21 dex, while galaxies with $M_{\star}$ values below the $\rm 25^{th}$-percentile have $\sim$10 Gyr and $\sim$0.15 dex. In this sense, we recover well-known relations between stellar population properties and stellar mass (see section \ref{intro}) at fixed $M_h$, as more massive galaxies are older and more metal-rich and than less massive ones. We also show this correlation between the scatter of the SHMR and the ages and [M/H] of the galaxies more quantitatively. We find that galaxies with the highest ages at fixed $M_h$ (red circles in the upper panel) have a significantly higher mean $M_{\star}$ than the ones with youngest ages (blue circles in the upper panel), especially at low halo masses. Very similar trends are found for metallicity (red and blue circles in the middle panel). In addition, we also observe how the lines of constant age and [M/H] across the SHMR are not horizontal, suggesting these parameters may not be only determined by stellar mass. We assess the dependence on different parameters more quantitatively in section~\ref{sec:par}, where we carry out a partial correlation analysis. Briefly, with this approach we are able to remove inter-correlations between the parameters and, hence, to truly evaluate the dependence between different parameters. In the bottom right corner of the panels we show the partial correlation coefficient strengths (solid black lines) between the stellar population parameters and $M_{\star}$ (vertical) and $M_h$ (horizontal). For reference, the grey solid lines have a length which corresponds to a (Spearman) correlation coeffcient of 0.5. The red solid line indicates the direction of maximal increase of the stellar population parameters. We find that the galaxy ages and [M/H] are also sensitive to halo mass, as these parameters also correlate with $M_h$. In addition, we observe how the direction of maximal increase of the stellar population properties is not vertical, meaning they are not only driven by stellar mass. These trends weaken at high halo masses, especially for the age, as galaxies are equally old and metal-rich regardless of $M_h$ and $M_{\star}$.With the partial correlation analysis, we also find that the correlations are milder in the high halo mass regime (see Fig.~\ref{fig:par_corr}). Our results are qualitatively in agreement with \citet{2021MNRAS.502.4457G}, who found that the ages and metallicites of centrals increase with stellar mass, and also that these parameters  do not depend on halo mass for high halo masses. \citet{2021MNRAS.500.4469T} also found that at fixed stellar mass, galaxies in less massive halos are more metal-rich, also in agreement with our findings. Although quantitatively there are slight discrepancies between our results, they might be mainly due to our different approaches to the problem and the difference in the halo mass estimation (which is discussed below). Given the complexity of how the stellar population properties are displayed over the stellar-to-halo mass relation as shown in Fig \ref{fig:smhm_pops},  we strongly advocate to use this relation instead of through one-dimensional projections of it.  

In Fig. \ref{fig:smhm_indices} we show the SHMR color-coded by the age-sensitive absorption index $\rm H\beta_o$ and the metallicity-sensitive indices $\rm [MgFe]^{\prime}$, Mgb5170 and $\rm <\!\!Fe\!\!>$, similarly to Fig. \ref{fig:smhm_pops}. In contrast to full-spectral fitting, this approach is model independent, yet we find that a remarkable agreement between the trends shown in both sets of figures.

\subsection{[Mg/Fe]} When considering $\alpha$ elements, scaling relations such as the well-known correlation between [$\alpha$/Fe] and stellar mass have been traditionally studied for early-type galaxies, as they are ideal candidates for for stellar population studies due to their reasonably simple star formation histories \citep{2005ApJ...621..673T}. In this work, we extend our analysis and measure [Mg/Fe] ratio for all galaxy types, which had not been done until recently \citep{2021MNRAS.502.4457G}. The bottom panel of Fig. \ref{fig:smhm_pops} shows that the [Mg/Fe] ratio seems to be decoupled from age and [M/H], as it does not follow the same trends across the SHMR shown in the upper and middle panels. Interestingly there seems to be a trend with halo mass, with  galaxies in more massive halos also being more Mg-enhanced.In this case, we see that more Mg-enhanced galaxies (red circles) tend to have a similar $M_{\star}$ than less Mg-enhanced galaxies (blue circles), and hence, the scatter in the SHMR does not seem to correlate significantly with the [Mg/Fe] ratio.

\section{Velocity dispersion - Halo mass relation}

\label{sec:sigmh}
\begin{figure}
    \centering
    \includegraphics[scale = 0.42]{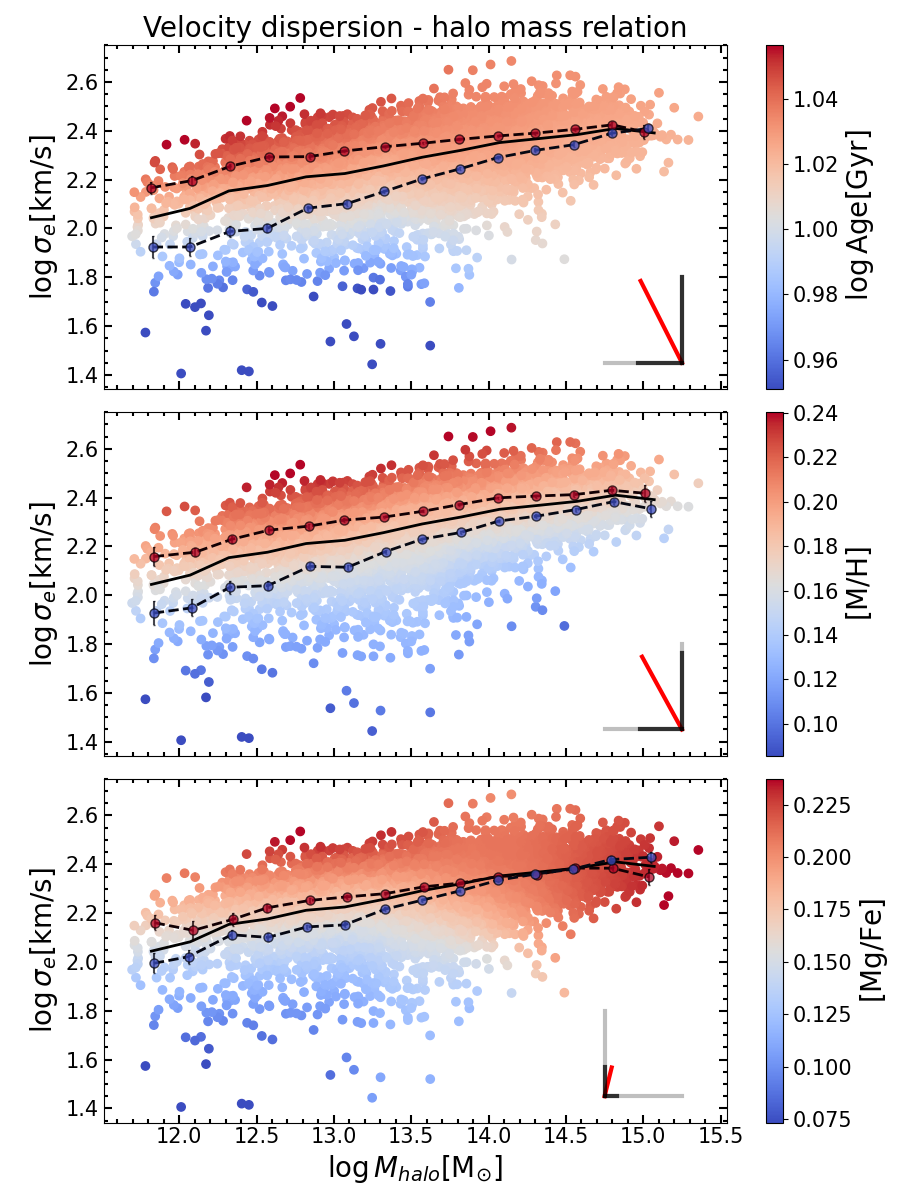}
    \caption{Velocity dispersion (measured within 1 $R_e$) as a function of halo mass for SDSS central galaxies color-coded by the stellar population properties resulting from our stellar population analysis described in section \ref{an_sp}: mass-weighted ages (upper panel), metallicities (middle panel) and [Mg/Fe] abundance (bottom panel).We applied the LOESS algorithm to the color-coding in order to highlight the global trends across the VDHMR. The median VDHMR is shown as a black solid line for reference. At fixed $M_h$, red and blue circles connected by dashed lines indicate the mean $\sigma_e$ of galaxies above the $\rm75^{th}$-percentile and below $\rm25^{th}$-percentile of the corresponding distributions. In the bottom right corner of the panels we show the partial correlation coefficient strengths (solid black lines) between the stellar population parameters and $\sigma_e$ (vertical) and $M_h$ (horizontal). For reference, the grey solid lines have a length which corresponds to a (Spearman) correlation coefficient of 0.5. The red solid line indicates the direction of maximal increase of the stellar population parameters (see section \ref{sec:par} for more details).  For low halo masses, the scatter correlates with the age and [M/H] at fixed $M_h$, while at fixed $\sigma_e$, galaxies in less massive halos are older and more metal-rich than the ones in more massive halos. At the high halo mass end, galaxies are mainly old and metal-rich. }
    \label{fig:sighm_pops}
\end{figure}

In addition to stellar mass, stellar population properties are known to correlate with velocity dispersion, showing the strongest and tightest correlations with $\sigma$ \citep[e.g.,][]{2009ApJ...698.1590G,2014MNRAS.445.1977L}. Hence, we also investigate how the stellar population parameters change across the velocity dispersion - halo mass relation. Fig. \ref{fig:sighm_pops} shows the velocity dispersion (measured within the SDSS fiber and corrected to 1 $R_e$ as detailed in section \ref{an_kine}) as a function of halo mass and color-coded by the mass-weighted ages (upper panel), metallicities (middle panel) and [Mg/Fe] abundance ratios (bottom panel) resulting from our stellar population analysis described in section \ref{an_sp}. To highlight the global trends of the stellar population parameters across the VDHMR, we applied the LOESS algorithm to the color-coding of Fig. \ref{fig:sighm_pops}. The median VDHMR is shown as a black solid line for reference. At fixed $M_h$, red and blue circles connected by dashed lines indicate the mean $\sigma_e$ of galaxies above the $\rm75^{th}$-percentile and below $\rm25^{th}$-percentile of the corresponding distributions.

 \subsection{Age and [M/H]} The trends of age and metallicity are similar to ones seen in Fig.~\ref{fig:smhm_pops} for the SHMR. Nevertheless, we observe a larger scatter at low halo masses with a transition from younger and more metal-poor galaxies with lower-$\sigma_e$ values to ones with higher-$\sigma_e$ that are older and more metal-rich. In the low halo mass regime, we observe that older galaxies at fixed $M_h$ (red circles in the upper panel) have a significantly higher mean $\sigma_e$ than younger ones (blue circles in the upper panel). We find very similar trends for [M/H] (red and blue circles in the middle panel). These differences at fixed $M_h$ are more noticeable in this plane than in the SHMR. We also note that the scatter of the VDHMR covers a wider range of ages and [M/H] than the SHMR, extending to younger ages and lower [M/H]. At $M_h$\,$\sim$\,10$^{12.5}$\,$\rm M_{\odot}$, the median age and [M/H] of the galaxies with $\sigma_e$ values above the $\rm 75^{th}$-percentile of  $\sigma_e$ is  $\sim$11 Gyr and $\sim$0.21 dex, while galaxies with  $\sigma_e$ values below the $\rm 25^{th}$-percentile have $\sim$9 Gyr and $\sim$0.1 dex. In this halo mass regime, $\sigma_e$ seems to be a better predictor of the ages and [M/H] than $M_{\star}$ (see also section \ref{sec:par}), as the lines of constant age or [M/H] are more horizontal compared to Fig. \ref{fig:smhm_pops}, but we also see how these lines are not completely horizontal, (i.e., they might not be independent of halo mass). To investigate this dependence, we also evaluate the partial correlations between the stellar population parameters and $\sigma_e$ and $M_h$ in section~\ref{sec:par}. Similarly to what we found for the SHMR, the direction of maximal increase of the stellar population properties is not vertical, as there is a significant correlation between ages and [M/H] and $M_h$. Therefore, we find that halo mass has a secondary yet relevant role regulating the observed ages and metallicities. As halo mass increases, the scatter decreases, and  we observe again that all the galaxies tend to be old for high halo masses. The correlation at fixed $M_h$ between [M/H] and the scatter of the SHMR seems weaker in this regime. We also find that the correlations are smaller when we repeat the partial correlation analysis only for the high halo mass regime (see Fig. \ref{fig:par_corr}).

We also show the age-sensitive index $\rm H\beta_o$ and the metallicity sensitive indices $\rm [MgFe]^{\prime}$, Mgb5170 and $\rm <\!\!Fe\!\!>$ across the VDHMR in Fig. \ref{fig:sighm_indices}. As for the SHMR, we also find that the indices show very similar trends to the ones seen for the age and [M/H] in Fig. \ref{fig:sighm_pops}. 

\subsection{[Mg/Fe]}Despite the fact that [Mg/Fe] also seems to differ from the other stellar population properties across the VDHMR, the trends are more similar than for the SHMR.We observe a trend with halo mass, with galaxies in more massive halos being more Mg-enhanced. Moreover, in this case we also see that the [Mg/Fe] abundance also seems to correlate with the scatter of the VDHMR for low halo masses, with more [Mg/Fe]-enhanced galaxies having higher $M_{\star}$ at fixed $M_h$.

\subsection{Robustness of the VDHMR} 
\begin{figure}
    \centering
    \includegraphics[scale = 0.42]{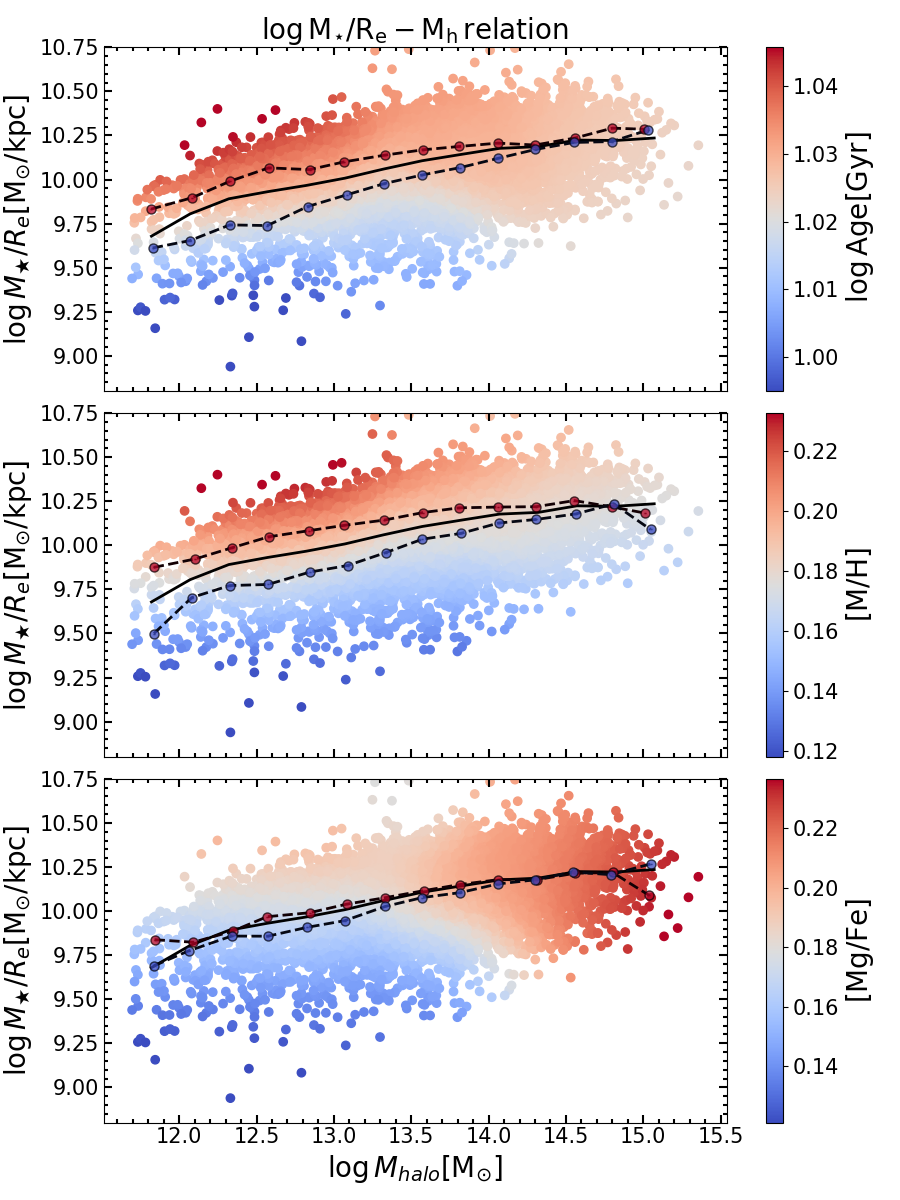}
    \caption{$M_{\star}/R_e$ as a function of halo mass for SDSS central galaxies color-coded by the stellar population properties resulting from our stellar population analysis described in section \ref{an_sp}: mass-weighted ages (upper panel), metallicities (middle panel) and [Mg/Fe] abundance (bottom panel).
   The median $M_{\star}/R_e-M_h$ relation is shown as a black solid line for reference. At fixed $M_h$, red and blue circles connected by dashed lines indicate the mean $M_{\star}/R_e$ of galaxies  with ages (upper panel), [M/H] (middle panel) and [Mg/Fe] abundances (bottom panel) above the $\rm75^{th}$-percentile and below $\rm25^{th}$-percentile of the corresponding distributions, respectively. Qualitative trends of the stellar population parameters across this relation are very similar to the ones across the VDHMR (Fig. \ref{fig:sighm_pops}).}
    \label{fig:sighm_pops_vir}
\end{figure}

While we have shown that both the SHMR and the VDHMR are useful tools in order to understand the  coupling between galaxies and their dark matter halos, we will focus on the VDHMR in the following sections. This choice is motivated by the fact that $\sigma_e$ is a better proxy for the stellar population properties (see also section~\ref{sec:par}). Moreover, both halo and stellar mass depend on the luminosity of the central galaxy, whereas the velocity dispersion is calculated independently.

Velocity dispersions are, however, not uniquely defined in numerical simulations. If we assume that the galaxies from our sample is virialized, $\log\sigma$ should be proportional to $\log M_{\star}/R_e$. To facilitate the comparison with numerical simulations, Fig. \ref{fig:sighm_pops_vir} shows the $\log M_{\star}/R$ - halo mass relation, color-coded by the stellar population properties, as in Fig. \ref{fig:sighm_pops}.  We note that the scatter in this relation is considerably lower than the one seen in the VDHMR. We observe how, in general, the qualitative trends of the stellar population properties across this relation have a similar behavior as the one seen for the VDHMR. The results for the [Mg/Fe] abundance are in agreement with the ones seen across the VDHMR. 

Our findings suggest that the VDHMR might be a good alternative to study these observed parameters. In this sense, the $\log M_{\star}/R_e - M_h$ relation might be an improvement with respect to the SHMR, as in addition to the stellar mass, the effective radius of the galaxies is also taken into account. We find that this relation could be used as proxy of the VDHMR and might be useful for future theoretical models and simulations.

\section{Dependence on halo mass}
\label{sec:par}
In sections \ref{sec:smhm} and \ref{sec:sigmh} we found that the ages and [M/H] of the galaxies not only depend on $M_{\star}$ or $\sigma_e$, but also depend on the masses of their host dark matter halos. In this section we assess more quantitatively the role of halo mass regulating the ages, [M/H] and [Mg/Fe] abundances. In this sense, as halo mass and the stellar population properties correlate with $M_{\star}$ and $\sigma_e$, we need to account for these dependencies in order to isolate the effect of halo mass.

\subsection{Residuals of scaling relations}

In this section we use the residuals of scaling relations in order to account for the stellar mass or velocity dispersion dependence.  We calculate the scatter in the scaling relations between the stellar population properties and $M_{\star}$ or $\sigma_e$ as the offset from the mean relation (the residuals). Then, we also obtain similarly the residuals of the halo mass - stellar mass relation. We study the correlations between these residuals to we asses whether there is also a secondary dependence of the stellar population properties on halo mass. 

\subsubsection{Removing the stellar mass dependence}
\begin{figure}
    \centering
    \includegraphics[scale = 0.48]{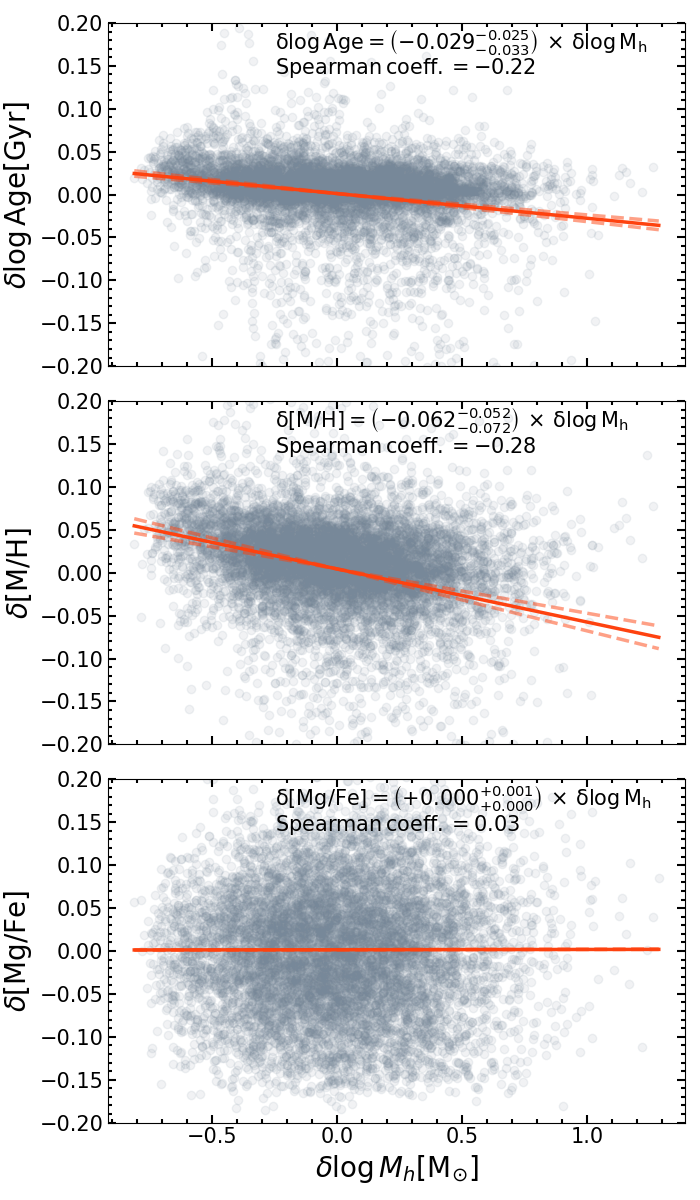}
    \caption{ Stellar population properties vs. halo mass for individual galaxies. We show the residuals of the logAge–$M_{\star}$ (upper panel), [M/H]–$M_{\star}$ (middle panel), [Mg/Fe]–$M_{\star}$ relation (bottom panel) plotted against the residuals of the $M_h$-$M_{\star}$ relation, for individual galaxies (grey circles). The solid lines correspond to the best-fitting relations and the dashed lines to the fit uncertainties. We also show the Spearman rank correlation coefficient. The residuals of the  logAge–$M_{\star}$ and [M/H]–$M_{\star}$ relations anticorrelate with the residuals of the $M_h$-$M_{\star}$ relation.}
    \label{fig:mscatter}
\end{figure}

First, we correct for the dependencies of the ages, [M/H] and [Mg/Fe] with $M_{\star}$. Fig. \ref{fig:mscatter} shows the residuals of the logAge–$M_{\star}$ (upper panel), [M/H]–$M_{\star}$ (middle panel), [Mg/Fe]–$M_{\star}$ relation (bottom panel) vs. the residuals of the $M_h$-$M_{\star}$ relation, for individual galaxies (grey circles). The solid lines correspond to the best-fitting relations and the dashed lines to the fit uncertainties. We also show the Spearman rank correlation coefficient. After accounting for dependence on $M_{\star}$, we observe that there are anticorrelations between the $M_h$ and the ages and [M/H] of the galaxies, with less massive halos hosting older and more metal-rich galaxies, at fixed $M_{\star}$. The correlation coefficients of the $\rm \delta\log Age$-$\delta M_h$ and $\rm \delta[M/H]$-$\delta M_h$ relations are -0.22 and -0.28, respectively, and we find the following the best-fitting relations:

\begin{equation}
    \rm \delta\log Age  = \left( {-0.029}^{-0.025}_{-0.033} \right) \, \times \, \delta \log \, M_h 
\end{equation}

\begin{equation}
    \rm \delta[M/H] = \left( {-0.062}^{-0.052}_{-0.072} \right) \, \times \, \delta \log \, M_h 
\end{equation}

Note that we do not show the normalization of these linear fits in these equations given that it is $\sim0$ by construction.

In contrast, the [Mg/Fe] ratio does not correlate with $M_h$ and we found a correlation coefficient of 0.03.

\subsubsection{Removing the velocity dispersion dependence}
\label{sec:rem_vd}
\begin{figure}
    \centering
    \includegraphics[scale = 0.48]{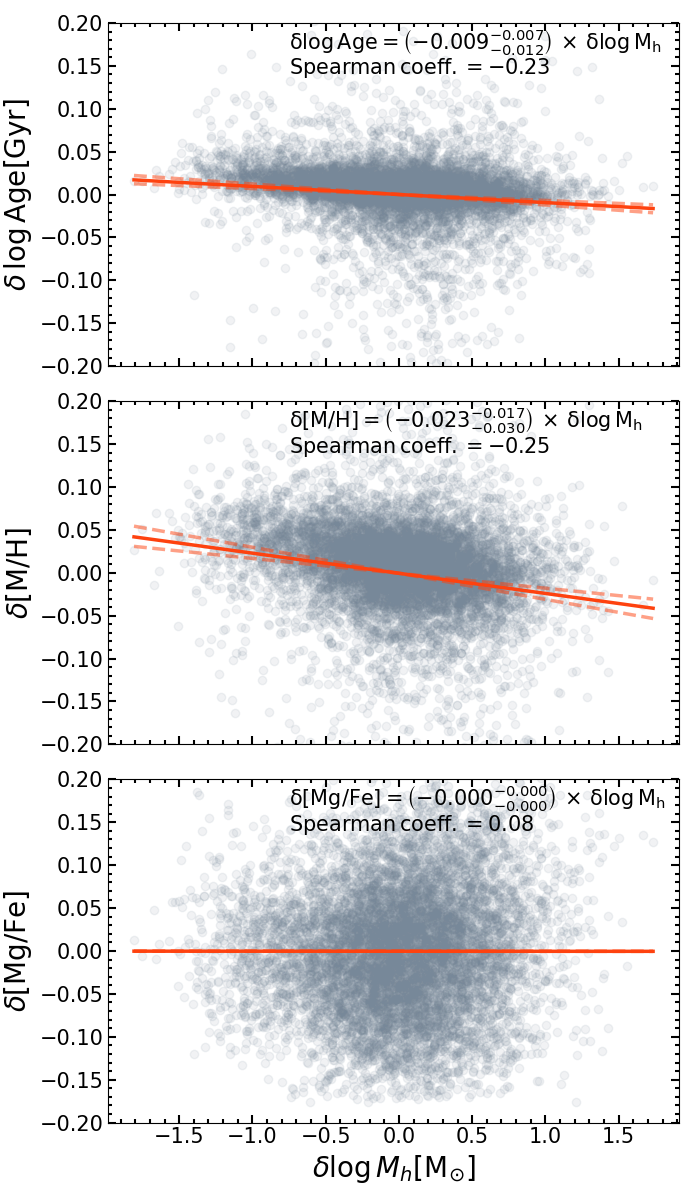}
    \caption{ Stellar population properties vs. halo mass for individual galaxies. We show the residuals of the logAge–$\sigma_e$ (upper panel), [M/H]–$\sigma_e$ (middle panel), [Mg/Fe]–$\sigma_e$ relation (bottom panel) plotted against the residuals of the $M_h$-$\sigma_e$ relation, for individual galaxies(grey circles). The solid lines correspond to the best-fitting relations and the dashed lines to the fit uncertainties. We also show the Spearman rank correlation coefficient. The residuals of the  logAge–$\sigma_e$ and [M/H]–$\sigma_e$ relations anticorrelate with the residuals of the $M_h$-$\sigma_e$ relation.}
    \label{fig:sigcatter}
\end{figure}

We repeat the previous analysis, but in this case we account for the dependencies of the ages, [M/H] and [Mg/Fe] with $\sigma_e$. In Fig. \ref{fig:sigcatter} we show the residuals of the logAge–$\sigma_e$ (upper panel), [M/H]–$\sigma_e$ (middle panel), [Mg/Fe]–$\sigma_e$ relation (bottom panel) vs. the residuals of the $M_h$-$\sigma_e$ relation, for individual galaxies (grey circles). The solid lines correspond to the best-fitting relations and the dashed lines to the fit uncertainties. We also show the Spearman rank correlation coefficient. After correcting for dependence on $\sigma_e$, we find anticorrelations between the $M_h$ and the ages and [M/H] of the galaxies, with less massive halos hosting older and more metal-rich galaxies. The correlation coefficients of the $\rm \delta\log Age$-$\delta M_h$ and $\rm \delta[M/H]$-$\delta M_h$ relations are -0.23 and -0.25, respectively, and the best-fitting relations are:

\begin{equation}
    \rm \delta\log Age  = \left( {-0.009}^{-0.007}_{-0.012} \right) \, \times \, \delta \log \, M_h 
\end{equation}

\begin{equation}
    \rm \delta[M/H] = \left( {-0.024}^{-0.017}_{-0.031} \right) \, \times \, \delta \log \, M_h 
\end{equation}

We observe that the halo mass dependence is smaller than in the previous case, (i.e, values of slopes are smaller). This is expected given that $\sigma_e$ is a better proxy of the stellar population properties than $M_{\star}$.  

In this case, the [Mg/Fe] abundance does not correlate with $M_h$ either, with a correlation coefficient of 0.08.

\subsection{Partial correlation analysis}
In addition, we also perform an alternative approach using partial correlation statistics. This method allows us to remove inter-correlations between the data in order to truly assess how stellar population properties depend on different parameters, in particular and most importantly for our study, halo mass. Specifically, the partial correlation coefficient \citep[][]{2017MNRAS.471.2687B,2020MNRAS.492...96B,2020MNRAS.499..230B} quantifies the strength of a correlation between two variables, keeping a third parameter fixed. This coefficient is defined as:

\begin{equation}
    \rho_{AB,C} = \frac{\rho_{AB} - \rho_{AC} \cdot \rho_{BC}}{\sqrt{1-\rho_{AB}^2}\sqrt{1-\rho_{BC}^2}}
    \label{eq:par_corr}
\end{equation}

which measures the correlation between variables A and B at fixed C, and where $\rho_{ij}$ is the Spearman rank correlation coefficients between parameters i and j. 

\begin{figure*}
    \centering
    \includegraphics[scale = 0.44]{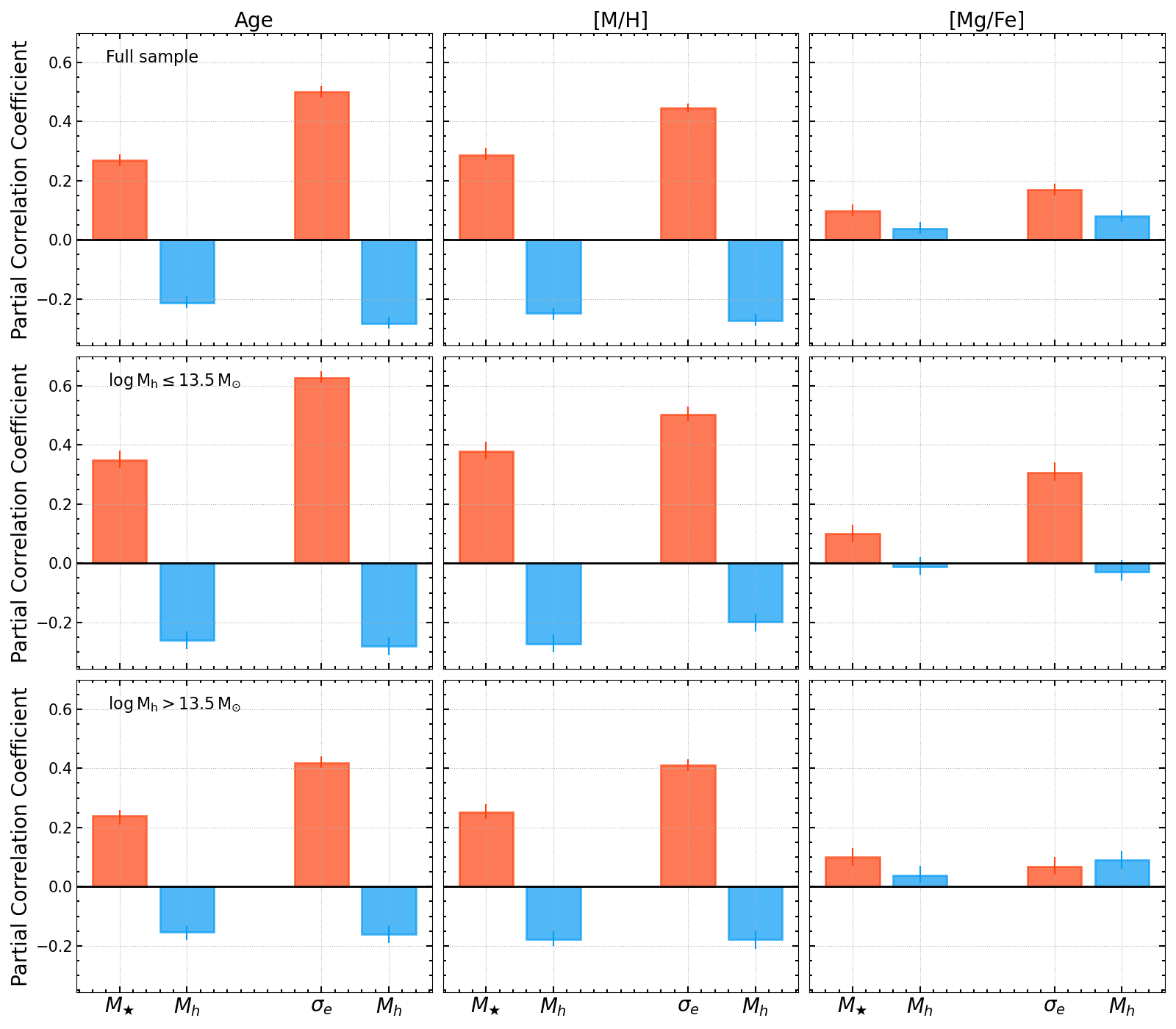}
    \caption{Stellar population parameters partial correlations. We show the partial correlation coefficient  between age (left column), [M/H] (middle column) and [Mg/Fe] (right column) and the parameters in the x-axis, which correspond from left-to-right to: $M_{\star}$ at fixed $M_h$,  $M_h$  at fixed $M_{\star}$, $\sigma_e$ at fixed $M_h$, and $M_h$ at fixed $\sigma_e$. We show the results for the whole sample (upper panels), for galaxies hosted by $M_h \leq 10^{13.5} M_{\odot}$ halos (middle panels), and for the ones in $M_h > 10^{13.5} M_{\odot}$ halos (bottom panels). The height of the bar indicates the  partial correlation coefficient strength.}
    \label{fig:par_corr}
\end{figure*}

Fig. \ref{fig:par_corr} shows the results of our partial correlation analysis. We show the partial correlation coefficient strength between age (left column), [M/H] (middle column) and [Mg/Fe] (right column) and the parameters in the x-axis, which correspond from left-to-right to: $M_{\star}$ at fixed $M_h$,  $M_h$  at fixed $M_{\star}$, $\sigma_e$ at fixed $M_h$, and $M_h$ at fixed $\sigma_e$. We show the results for the whole sample (upper panels), for galaxies hosted by $M_h \leq 10^{13.5} M_{\odot}$ halos (middle panels), and for the ones in $M_h > 10^{13.5} M_{\odot}$ halos (bottom panels). The height of the bar indicates the strength partial correlation coefficient. 

First, focusing on the full sample (upper panels), we find very similar results for age (left panel) and [M/H] (middle panel), as expected given the similar trends found for these parameters across the SHMR and across the VDHMR. With this method we recover through more quantitative approach the results observed visually in Figs. \ref{fig:smhm_pops} and \ref{fig:sighm_pops}. In addition to the expected correlation between age ([M/H]) and $M_{\star}$ at fixed $M_h$, we also find a secondary anticorrelation between age ([M/H]) and $M_h$, once the dependence with $M_{\star}$ is removed (in agreement with Fig. \ref{fig:mscatter}). Moreover, we observe that the age and [M/H] correlate more strongly with $\sigma_e$, However, when this dependence is removed, we also observe an anticorrelation between the age ([M/H]) and $M_h$. On the other hand, we observe significantly milder correlations for the [Mg/Fe] abundance. In this case, [Mg/Fe] correlates weakly with $M_{\star}$ (at fixed $M_h$), and even weaker with $M_h$ (at fixed $M_{\star}$). Similarly to the other parameters, the correlation with $\sigma_e$ (at fixed $M_h$) is the strongest, although still very mild. There is a very small secondary correlation found with $M_h$ once the dependence with $\sigma_e$ is removed. 

We show these correlation strengths as solid black lines in the bottom right corner of Figs. \ref{fig:smhm_pops} and \ref{fig:sighm_pops}. For reference, the length of the grey solid lines correspond to a correlation strength of 0.5. Following a similar approach to the one employed in \citet{2020MNRAS.492...96B}, we also compute the direction of maximal increase of the stellar population parameters across the SHMR and the VDHMR. The slope of the axis is given by the ratio of the partial correlation coefficients $\rho_{YZ,X}$ and $\rho_{XZ,Y}$, where X corresponds to $M_h$, Z corresponds to the stellar population parameter (age, [M/H] or [Mg/Fe]), and Y to $M_{\star}$ in the SHMR and to $\sigma_e$ in the VDHMR. We show these axes as a red solid lines in the bottom right corner of the panels of Figs. \ref{fig:smhm_pops} and \ref{fig:sighm_pops}.

Moreover, in sections \ref{sec:smhm} and \ref{sec:sigmh} we found that the trends of ages and [M/H] across the SHMR and VDHMR seem to differ above and below $M_h=10^{13.5} \, \rm M_{\odot}$. For that reason, we also repeat the partial correlation analysis for these two regimes (middle and bottom panels of Fig. \ref{fig:par_corr}). As we already observed in Figs.  \ref{fig:smhm_pops} and \ref{fig:sighm_pops}, for age and [M/H] the correlations are stronger at $M_h \leq 10^{13.5}\, \rm M_{\odot}$ than  at $M_h > 10^{13.5}\, \rm M_{\odot}$, and also stronger than for the full sample. The behavior of the [Mg/Fe] is less clear. The correlations at $M_h > 10^{13.5}\, \rm M_{\odot}$ are very mild and similar to the ones for the whole sample, except for $\sigma_e$ (at fixed $M_h$) which is considerably smaller. However, the correlation strength $\sigma_e$ (at fixed $M_h$) is significantly larger at $M_h \leq 10^{13.5}\, \rm M_{\odot}$. We observe no significant correlations with $M_h$ (at fixed $\sigma_e$ or at fixed $M_{\star}$).

\section{Dependence on morphology}
\label{sec:morph}
To assess the possible effect of morphology on the stellar population properties across the VDHMR, we divide the galaxies into early- and late-type galaxies using the classification of \cite{2018MNRAS.476.3661D}. We explore the VDHMR for early- and late-type galaxies separately, and hence we are able study the trends for early-types only, as traditionally done in stellar population studies \citep[e.g.,][]{1973ApJ...179..731F,1989PhDT.......149P,1992ApJ...398...69W,2000AJ....119.1645T,2005MNRAS.362...41G,2005ApJ...621..673T,2015MNRAS.448.3484M,2018MNRAS.475.3700M}.

\begin{figure*}
    \centering
    \includegraphics[scale = 0.45]{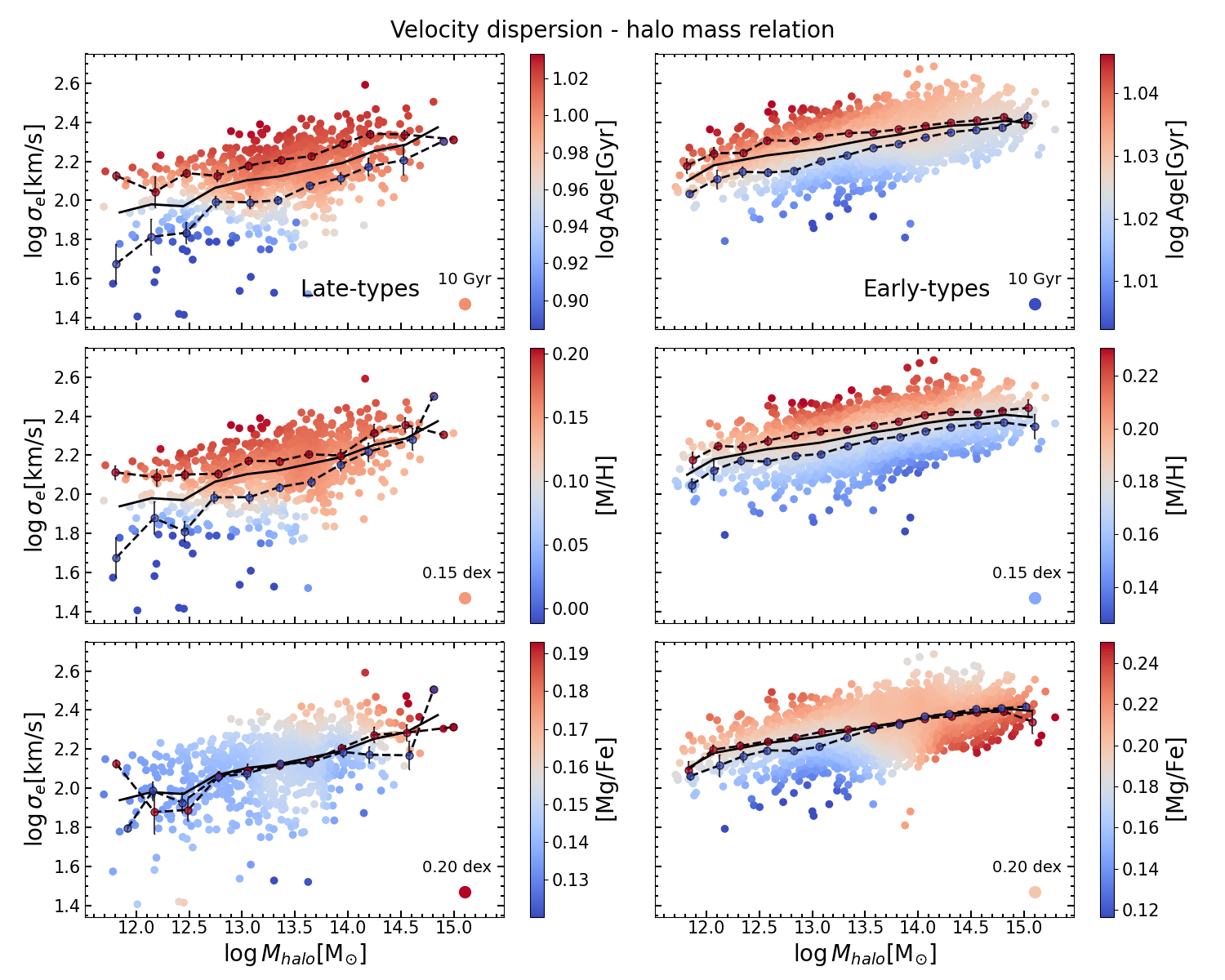}
    \caption{The dependence of the velocity dispersion - halo mass relation on morphology. Here, we show the VDHMR separately for for late- and early-type galaxies on the left and right columns, respectively. Analogously to Figure \ref{fig:sighm_pops}, the VDHMR is color-coded by the stellar population properties resulting from our stellar population analysis described in section \ref{an_sp}: mass-weighted ages (upper panels), [M/H] (middle panels) and [Mg/Fe] (lower panels). For reference,  there is a circle in the bottom right corner indicating the color of a galaxy with an age of 10 Gyr (upper panels), metallicity of 0.15 dex (middle panels) and [Mg/Fe] abundance of 0.20 dex. We applied the LOESS algorithm to the color-coding in order to highlight the global trends across the VDHMR. The median VDHMR is shown as a black solid line for reference. At fixed $M_h$, red and blue circles connected by dashed lines indicate the mean $\sigma_e$ of galaxies above the $\rm75^{th}$-percentile and below $\rm25^{th}$-percentile of the corresponding distributions.  Early types have considerably less scatter across the VDHMR than late-types at the low mass end. In this regime, late-types extend to lower-$\sigma_e$, younger ages and lower [M/H] than early-types, as the latter concentrate in the high-$\sigma_e$ region with older ages and higher metallicites.}
    \label{fig:sighm_pops_morph}
\end{figure*}

Fig. \ref{fig:sighm_pops_morph} shows the VDHMR for late-type (left columns) and early-type galaxies (right columns), separately. The columns of this figure are analogous to the ones of Fig. \ref{fig:sighm_pops}, but we also include a reference in the bottom right corner indicating the color of a galaxy with an age of 10 Gyr (upper panels), metallicity of 0.15 dex (middle panels) and [Mg/Fe] abundance of 0.20 dex. Note that the reference circles tend to have bluer colors than in the case of halos in early-types than late-types. Therefore, early-types are on average older, more metal-rich and more Mg-enhanced than late-types.

\subsection{Late-type galaxies} 

 Compared to Fig. \ref{fig:sighm_pops}, we see how the shape of the VDHMR for late-types (left columns of Fig. \ref{fig:sighm_pops_morph}) is slightly different mainly due to the considerably low number of late-type galaxies in our sample. In this sense, we observe that late-types tend to populate the low halo mass end with lower $\sigma_e$ values  than early-types (right column). These galaxies have an average halo mass of $10^{13.4} \, \rm M_{\odot}$ and an average $\sigma_e$ of $\sim$133 km/s.

\paragraph*{Age and [M/H]} The results for age (upper left panel) panel and [M/H] (middle left panel) across this relation are very similar to the ones shown in Fig. \ref{fig:sighm_pops}. At low halo masses, we observe again considerably scatter in the relation and despite the low number of late-types, we still see the transition from young and metal-poor galaxies to old and metal rich ones when increasing $\sigma_e$.In this regime and at fixed $M_h$, we generally find that galaxies with the oldest ages and highest [M/H] (red circles) are above the median VDHMR and have considerably higher $\sigma_e$ than the youngest and more metal-poor galaxies (blue circles), which are below the median VDHMR. Late-types in very massive halos seem to be mainly older and more metal-rich as well, but we note the lack of them at the very massive end. Compared to early-types, the scatter is larger at low halo masses. Late-types extend to lower $\sigma_e$ values, younger ages and lower [M/H], as expected for this type of galaxies \citep[e.g.,][]{2005MNRAS.362...41G,2015A&A...581A.103G}. A  $M_h$\,$\sim$\,10$^{12.5}$\,$\rm M_{\odot}$, the median age and [M/H] of the galaxies with $\sigma_e$ values above the $\rm 75^{th}$-percentile of  $\sigma_e$ is $\sim$9.4 Gyr and $\sim$0.16 dex, while galaxies with  $\sigma_e$ values below the $\rm 25^{th}$-percentile have $\sim$6.5 Gyr and $\sim$-0.11 dex.

\paragraph*{[Mg/Fe]} In this case, the [Mg/Fe] abundance (bottom left panel of  Fig. \ref{fig:sighm_pops_morph}) does not seem to follow the trends of the other stellar population parameters.  We do not observe any clear trend, and the more Mg-enhanced galaxies at fixed $M_h$ (red circles) seem to have similar $\sigma_e$ values than the galaxies with the lowest [Mg/Fe] abundances (blue circles). However, we observe again a very mild trend with $M_h$, with galaxies in very massive halos being more Mg-enhanced.\\

In the left panels of Fig. \ref{fig:smhm_pops_morph} we show the same morphological analysis but across the SHMR. The results for the age and [M/H] are in agreement with the ones found for the VDHMR, yet the results for the [Mg/Fe] remain uncertain. 

\subsection{Early-type galaxies}
\label{sec:morph_early}
The right panels of Fig. \ref{fig:sighm_pops_morph} show the VDHMR for early-types. Regarding the shape of the relation, the scatter significantly decreases at the low mass end in comparison to  Fig. \ref{fig:sighm_pops}. At  $M_h$\,$\sim$\,10$^{12.5}$\,$\rm M_{\odot}$, the standard deviation of the $\log \sigma_e$ distribution decreases from 0.15 dex (whole sample) to 0.1 dex (early-types only). These galaxies tend to populate the high-$\sigma_e$ region of the relation, and also the high halo mass end, as opposed to late-types. These early-types have an average $\sigma_e$ of 211 km/s and an average halo mass of $10^{13.7} \, \rm M_{\odot}$.

\paragraph*{Age and [M/H]}  We observe a transition from  younger and metal-poor galaxies to older and metal-rich ones for galaxies in low-mass halos similar to the the findings for the main sample, but with significantly less scatter and with these galaxies concentrating in the region of high $\sigma_e$, old ages and high metallicities.  At $M_h$\,$\sim$\,10$^{12.5}$\,$\rm M_{\odot}$, the median age and [M/H] of the galaxies with $\sigma_e$ values above the $\rm 75^{th}$-percentile of  $\sigma_e$ is $\sim$11 Gyr and $\sim$0.21 dex, while galaxies with  $\sigma_e$ values below the $\rm 25^{th}$-percentile have $\sim$10.5 Gyr and $\sim$0.16 dex. Here, we also recover that for a given $M_h$ galaxies with the highest ages and [M/H] (red circles) have higher $\sigma_e$ than galaxies with the lowest values of age and [M/H]. In very massive halos, we observe again that galaxies are mainly old. In this regime, the correlation of [M/H] with the scatter of the relation does not seem to weaken as much as for age.  The majority of the galaxies are early-types, and hence the trends are very similar to the ones seen for the whole galaxy population, although the range of age and [M/H] covered is narrower than for the whole sample. To sum up, we generally see that early-types correspond to the old and metal-rich galaxies seen in the upper region of the VDHMR (upper and middle panels of Fig. \ref{fig:sighm_pops}), as expected for this type of galaxies.

\paragraph*{[Mg/Fe]} Regarding the [Mg/Fe] abundance (bottom right panel of  Fig. \ref{fig:sighm_pops_morph}), its behavior is quite similar to the one of the main sample (Fig. \ref{fig:sighm_pops}). At low halo masses and at fixed $M_h$, the scatter seems to correlate with the [Mg/Fe] abundance. At the same time, we observe a  trend with halo mass with galaxies in more massive halos being also more Mg-enhanced. \\

In Appendix \ref{ap:morph} we also study the SHMR for early-types (right panels of Fig. \ref{fig:smhm_pops_morph}). We find that the trends for age, [M/H] and [Mg/Fe] are in agreement with the results found for the VDHMR in the high halo mass regime.  

\section{Dependence on environment}
\label{sec:env}

\begin{figure*}
    \centering
    \includegraphics[scale = 0.45]{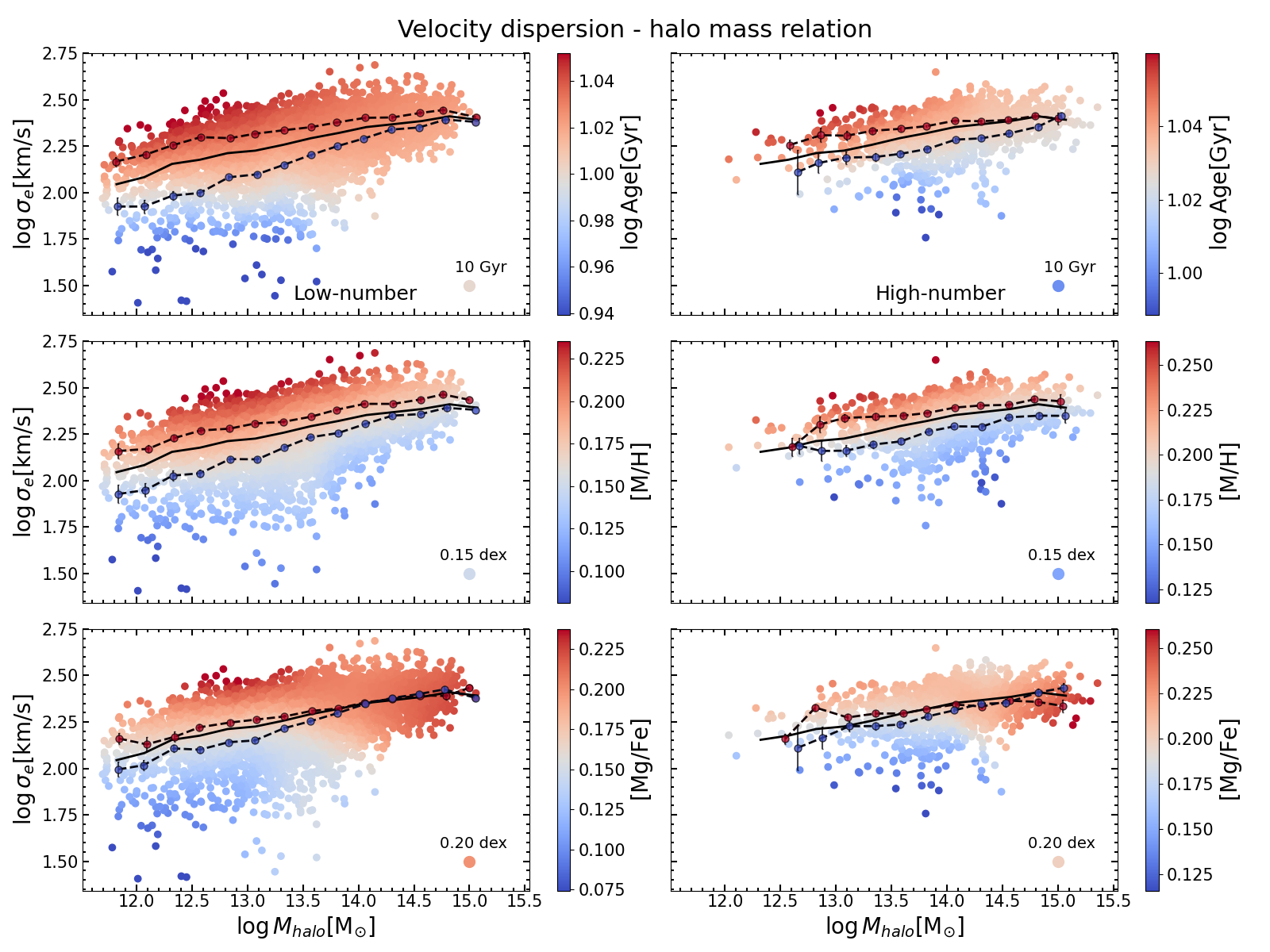}
    \caption{The dependence of the velocity dispersion - halo mass relation on environment. 
    Here, we show the VDHMR for galaxies in different environments, which correspond to different columns: galaxies in halos with low- (left column) and high-number of galaxies (right column). Similar to Fig. \ref{fig:sighm_pops}, the VDHMR is color-coded by the stellar population properties resulting from our stellar population analysis described in section \ref{an_sp}: mass-weighted ages (upper panels), [M/H] (middle panels) and [Mg/Fe] (lower panels). For reference,  there is a circle in the bottom right corner indicating the color of a galaxy with an age of 10 Gyr (upper panels), metallicity of 0.15 dex (middle panels) and [Mg/Fe] abundance of 0.20 dex. We applied the LOESS algorithm to the color-coding in order to highlight the global trends across the VDHMR. The median VDHMR is shown as a black solid line for reference. At fixed $M_h$, red and blue circles connected by dashed lines indicate the mean $\sigma_e$ of galaxies above the $\rm75^{th}$-percentile and below $\rm25^{th}$-percentile of the corresponding distributions. The trends across the VDHMR for galaxies in halos where there is a low number of galaxies are very similar to the ones seen in Fig. \ref{fig:sighm_pops} (except for high halo masses). For galaxies in halos with high number of galaxies, they are also quite similar, although the galaxies tend to be older  and more metal-rich, similar to what is found for early-types (Fig. \ref{fig:sighm_pops}).}
    \label{fig:sighm_pops_env}
\end{figure*}

There is observational evidence suggesting that the environment in which galaxies form can influence their evolution and leave an imprint in galaxy properties \citep[e.g.,][]{1980ApJ...236..351D,2003ApJ...584..210G,2004MNRAS.348.1355B,2006MNRAS.373..469B,2006MNRAS.366....2W,2008ApJ...684..888P,2008MNRAS.390..245C,2010MNRAS.407..937P,2010ApJ...721..193P,2013MNRAS.428.3306W,2013MNRAS.436...34C,2021MNRAS.502.4457G,2021MNRAS.500.4469T}. We also investigate how the stellar population properties of central galaxies across the VDHMR can be affected by their environment. We note that there are many different approaches to characterize galaxy environment and it is conceptually complex to define it. For simplicity, we use the number of galaxies within each halo as a metric to characterize it. We divide our sample of central galaxies into two subsamples using the number of galaxies in the halo (including the central), $N_{gal}$: halos with low- ($3 < N_{gal}< 10$) and high-number of galaxies ($N_{gal}\geq10$), with 6,977 and 1,824 galaxies in each category, respectively.

The columns of Fig. \ref{fig:sighm_pops_env} show the VDHMR for central galaxies within halos hosting low- (left) and high- (right) number of galaxies, analogously to Fig. \ref{fig:sighm_pops}. In order to facilitate the visualization of the trends, there is a circle for reference in the bottom right corner indicating the color of a galaxy with an age of 10 Gyr (upper panels), metallicity of 0.15 dex (middle panels) and [Mg/Fe] abundance of 0.20 dex.

\subsection{Low-number of galaxies} We show the VDHMR for central galaxies in halos with low number of galaxies in the left panels Fig. \ref{fig:sighm_pops_env}, which is analogous to Fig. \ref{fig:sighm_pops}, but only for centrals of halos with $3<N_{gal}<10$. The shape of the VDHMR is similar to one observed in  Fig. \ref{fig:sighm_pops}, except for the lack of galaxies in very massive halos end. The trends of the stellar population parameters across this relation are almost identical to the results shown in Fig. \ref{fig:sighm_pops}, as the majority of the objects reside in halos with low number of galaxies. 

\subsection{High-number of galaxies} In the right panels of Fig. \ref{fig:sighm_pops_env} we show the VDHMR relation for galaxies of halos with  $N_{gal}\geq10$. The number of centrals in this type of halos is considerably low, and galaxies tend to populate the high-$\sigma_e$ of the relation and very massive halos. Similarly to what we found for early-types, at fixed halo mass, these galaxies have a narrower range of age (upper panel) and [M/H] (middle panel), being generally older and more metal-rich compared to Fig. \ref{fig:sighm_pops}. We note that the reference circle, which indicates the color of the galaxies with an average age of 10 Gyr (upper panel), [M/H] of 0.15 dex (middle panel) and [Mg/Fe] of 0.20 dex (bottom panel), tend to have bluer colors than in the case of halos with less number of galaxies (for age and [M/H]). This means that, on average, central galaxies become progressively older, more metal-rich in halos with increasing number of satellites. We observe very similar trends to the ones of the whole sample, although we take into account that they are less defined given the lack of galaxies in the low-$\sigma_e$ region and the low mass end. For a given halo mass, we still observe a transition from younger and more metal-poor galaxies, to older and more metal-rich ones and that galaxies in very massive halos tend to have similar ages. The trends of the [Mg/Fe] abundance are also very similar to the ones of the whole sample. 

\paragraph*{} Our results for galaxies residing in halos with high-number of galaxies have similarities with the ones found for early-type galaxies in section \ref{sec:morph_early}. For both early-types and galaxies in these environments, we found that the shape of the VDHMR is different to the one shown in Fig. \ref{fig:sighm_pops}, with galaxies concentrating in the high-$\sigma_e$ region of the relation. The similarity between these trends seems to be in agreement with the morphology-density relation \citep[e.g.,][]{1980ApJ...236..351D,2003MNRAS.346..601G,2011MNRAS.416.1680C,2017ApJ...851L..33G,2017ApJ...844...59B}. This relation indicates that early-type galaxies are preferentially located in high-density environments, while late-types in low-density ones, with the fraction of early-types increasing with galaxy density.

\section{Discussion}
\label{sec:discussion}

\subsection{Physical origin}
In sections \ref{sec:smhm} and \ref{sec:sigmh} we found a correlation  of the scatter of the SHMR and the VDHMR with the ages, metallicities and [Mg/Fe] of the galaxies (in particular, for low halo masses, $M_h < 10^{13.2} M_{\odot}$). In this section, we discuss the potential drivers of these observed trends, and in particular, their possible relation with halo formation times and velocity dispersion of the galaxies. 

\subsubsection{Halo formation time}
Theoretical studies have found that the scatter of the SHMR correlates with different halo properties at fixed $M_h$, such as the concentration and the halo formation time (see section \ref{intro}). However, it is not straight-forward to connect observed galaxy properties, such as the ages and [M/H] of galaxies, with different halo properties, as different theoretical models differ in their predictions. 
 
 On one hand, a correlation between halo formation time and the scatter of the SHMR is found when using different hydro-simulations and semi-analytical models: at fixed halo mass, earlier-formed halos host more massive galaxies than late-formed ones \citep[e.g.,][]{2013MNRAS.431..600W,2017MNRAS.465.2381M,2017MNRAS.470.3720T,2018ApJ...853...84Z,2018MNRAS.480.3978A,2019MNRAS.490.5693B,2020MNRAS.491.5747M,2021arXiv210512145C}. Moreover, following the work of  \cite{2017MNRAS.465.2381M} with hydrodynamical simulations from the EAGLE project \citep{2015MNRAS.446..521S}, at fixed $M_h$  \citet{2019MNRAS.484..915M} found that galaxies hosted by earlier-formed halos had a higher SFR in the past and end up with  higher $M_{\star}$, but lower SFRs today than the ones formed within late-formed halos. 
 
 Regarding our findings, we find evidence that there is a connection between the properties of dark matter halos and the stellar population properties of nearby galaxies. We observe that halo mass has a role driving the ages and [M/H] of the galaxies, in addition to $M_{\star}$ and $\sigma_e$. At the same time, we find that older and more-metal rich galaxies have higher $M_{\star}$ and $\sigma_e$ at fixed $M_h$, suggesting that they have been more efficient in forming stars (Fig. \ref{fig:smhm_pops} and Fig. \ref{fig:sighm_pops}). In this regard, we found that galaxies in the upper region of the SHMR and the VDHMR tend to have old and metal-rich stellar populations at fixed halo mass. While theoretically, these massive galaxies (in the upper region of the SHMR at fixed halo mass) are hosted by halos that assembled earlier than halos hosting less massive galaxies \citep[e.g.,][]{2017MNRAS.465.2381M}. Hence, we speculate that older and more metal-rich galaxies reside within halos that formed earlier.  In this picture, early-formed halos host galaxies which  have had high SFR in the past \citep[e.g.,][]{2019MNRAS.484..915M}, assembled early, quenched early, and are chemically more evolved, leading to the stellar population properties observed at $\rm z=0$ (see Figs. \ref{fig:smhm_pops} and \ref{fig:sighm_pops}).  

We note that a different scenario is portrayed using abundance matching models. \cite{2013ApJ...770...57B} found that for low mass halos ($M_h \leq 10^{12}$ $\rm M_{\odot}$) the efficiency of converting baryons into stars is higher at present day than at high redshift. This suggests that halos that formed later are more efficient in forming stars than earlier-forming ones of the same mass, as they have spend more time in the regime of more efficient star formation. Nevertheless, we note that the regions of the SHMR and the VDHMR populated by galaxies at $z=0$ are not necessarily the same regions that were populated when they formed. In this regard, the star formation efficiency observed today does not have to be the same as the one at birth. For example, galaxies dominated by old stellar populations, which have had a prolonged evolution, might had experienced mergers, accreting more or less baryions than dark matter. This can lead to a different ratio of stellar mass and halo mass, and hence, change the star formation efficiency over time.

\subsubsection{Velocity dispersion}
Whether either stellar mass or velocity dispersion is most fundamental to determine the properties of galaxies and their host halos, it is still under debate \citep[e.g.,][]{2013ApJ...762L...7L,2016ApJ...832..203Z}, given that these two parameters strongly correlate with each other \citep[e.g.,][]{2016ApJ...832..203Z}. However, different authors suggest that $\sigma$ 
is a better observable to connect galaxies to their dark matter halos \citep[e.g.,][]{2012arXiv1201.1913W,2015ApJ...800..124B,2016ApJ...832..203Z}. In addition, scaling relations of stellar population properties with $\sigma$ are tighter. We find that when studying the dependence of the stellar population properties on halo mass, the regions of constant age or [M/H] across the VDHMR, are more horizontal than the ones across the SHMR, and that these properties correlate most strongly with $\sigma_e$, indicating that $\sigma_e$ is a better predictor of the stellar population properties than $M_{\star}$.  

 In the $\Lambda$CDM paradigm, halo mass is thought to be the main driver of the growth and evolution of galaxies \citep[e.g.,][]{white1978core,1984Natur.311..517B,2013ApJ...770...57B}. We observe indeed that halo mass has a role regulating the stellar population properties, but it is secondary, with $\sigma_e$ mainly driving the trends of stellar population properties across the VDHMR (Fig. \ref{fig:sighm_pops}). The velocity dispersion trace the stellar kinematics and it is related to the gravitational potential of the galaxies. In particular, \citet{2018ApJ...856...64B} suggested that the stellar population properties are ultimately regulated by this gravitational potential, as galaxies with deeper potential wells and higher escape velocities are more capable of retaining their stellar ejecta. On the other hand, $\sigma$ also contains information about the super-massive black holes in their centers, given the empirical correlation between black hole mass and $\sigma$ \citep[e.g.,][]{2000ApJ...539L...9F,2000ApJ...539L..13G,2009ApJ...698..198G,2012MNRAS.419.2497B,2016ApJ...831..134V}. Assuming that galaxies are virialized systems, this implies a relation between the velocity dispersion, the gravitational potential and the mass of the super-massive black hole. However, for massive galaxies, we also note that black hole feedback is expected to be the dominant effect, as for example, the stellar population properties show variations with the black hole mass at a given $\sigma$ \citep{2016ApJ...832L..11M,2018Natur.553..307M,2020ApJ...898...83D}, and hence for a given gravitational potential. In this respect, we argue that the trends that we see in the stellar population properties across the SHMR and the VDHMR might be produced as a combination of how super-massive black holes affect their host galaxies  \citep[e.g.,][]{2018Natur.553..307M} and halo properties. \\
 
Finally, it is worth noting that spectroscopy is required to obtain velocity dispersion measurements, which are observationally more expensive than photometric stellar masses. In addition, velocity dispersion is not necessarily  defined in numerical simulations in a manner that is fully consistent with observations due to projection and numerical effects. Hence, the $M_{\star}/R_e$ ratio could be used as an approximate metric, as shown in Fig. \ref{fig:sighm_pops_vir}. Alternatively, the central stellar surface mass density has also been proposed as a good predictor of the stellar population properties \citep[e.g.,][]{2012ApJ...760..131C,2013ApJ...776...63F}. This structural quantity, which is well correlated with velocity dispersion, traces bulge growth and is an indicator of quenching.

\subsection{Morphology}

\cite{2020MNRAS.499.3578C} also analyzed how the SHMR is affected by the morphology of the galaxies analyzing SDSS galaxies from the \citet{2007ApJ...671..153Y} catalog. At the low halo mass end and fixed $M_h$ ($M_h<12.9$), they find that disc galaxies are more massive than ellipticals. This seems to be in disagreement with our results, as we find that old and metal-rich galaxies are the most massive. These are typical characteristics of early-type galaxies, and we also find that this high-mass region tend to be dominated early-types. Nevertheless, the trend found by these authors disappear when they do tests using another $M_{\star}$ estimates. We also note that as opposed to our study, they used the halo mass estimate based on the $M_{\star}$ of the groups, $M_h(M_c)$.
\cite{2020MNRAS.499.3578C} also studied the scatter of the SHMR for central  galaxies from EAGLE and found that for low halo masses, earlier-formed halos host galaxies with more disc-like features. As the host halos of these discs formed earlier, these authors propose that these galaxies had more time for star formation and accretion. At first, this does not seem to be in line with our findings, as discs tend to be associated with younger stellar populations. Nonetheless, we note that we follow rather different approaches. \cite{2020MNRAS.499.3578C} used the fraction of kinetic energy in ordered co-rotation, $\kappa_{co}$, (which they found to correlate with the scatter of the SHMR at low halo masses) as an indicator of disc-like morphology. They found that galaxies with higher $\kappa_{co}$ (more disc-like) formed earlier and have higher $M_{\star}$ (at fixed $M_h$). In this scenario, galaxies in the lower region of the SHMR, which we find to be younger and more metal-poor, would be less supported by rotation than older more metal-rich galaxies (that have higher $M_{\star}$ and $\sigma_e$). We note, however, that as the SDSS fiber has a diameter of 3'', we measure $\sigma$ only in the inner region of the galaxies, while $\kappa_{co}$ is computed using a larger extent. Hence, at the low halo mass regime, if we measured $\sigma$ to a larger extent, we could also have lenticular galaxies both with high $\sigma$ and high rotation.

\subsection{Colors}
\label{sec:disc_color}

At the moment there is no agreement between different studies that investigate the behavior of the SHMR for red (or passive) and blue (or star-forming) galaxies \citep[see section 6.1 of ][]{2018ARA&A..56..435W}. At fixed $M_h$,  different authors find that red (or quiescent) galaxies reside within either more massive halos, less massive halos, or within halos of similar mass than blue (or star-forming) galaxies. Although we note that they use different analysis and techniques to estimate halo masses, in addition to different stellar mass estimates. 

In our study, we observe that older and more-metal rich galaxies, populations that generally have redder colors, have higher $M_{\star}$ and $\sigma_e$ at fixed $M_h$, and in particular at low halo masses. Our results are in agreement with \cite{2016MNRAS.455..499L}. They analyzed central SDSS galaxies from \citet{2007ApJ...671..153Y} and found that galaxies with higher $M_{\star}/M_h$ are redder, more quenched in star formation, and more bulge-dominated. In contrast, observationally, \citet{2016MNRAS.457.3200M} and \cite{2020MNRAS.499.3578C} found that at fixed $M_{\star}$ red galaxies reside in more-massive halos than blue galaxies. Both studies use the same stellar mass estimate from \citet{2003MNRAS.341...33K}, but \citet{2016MNRAS.457.3200M} used halo mass estimates obtained through galaxy-galaxy lensing. These results also seem to be in disagreement ours, as red galaxies are typically associated with old stellar populations, which we find to be in less massive halos at fixed $M_h$ (for low halo masses). Nevertheless, we note again the difference in the $M_{\star}$ calculation and the different choice of the halo mass estimates. Using the hydro-simulation SIMBA \citep{2019MNRAS.486.2827D},  \citet{2021arXiv210512145C} found a similar trend  as the one found by \citet{2016MNRAS.457.3200M} and \cite{2020MNRAS.499.3578C}. However, it is not observed when \cite{2020MNRAS.499.3578C} repeat the analysis for simulated galaxies from EAGLE.

There must be also taken into account that inferring the galaxy properties such as ages and metallicities with detailed stellar population analyses leads to a considerable improvement with respect to the ones derived with photometric colors, given that the colors of old populations suffer from the age-metallicity degeneracy \citep[e.g.,][]{1994ApJS...95..107W,2011MNRAS.415..709S}. Although our stellar population measurements, and hence our mass growth determination, are an advancement with respect to color-based measurements, we note that there are still systematic errors both in the stellar population analysis and halo mass determination. Nevertheless, we are confident in our stellar populations results, given that they follow well-known scaling relations and are in agreement with model-indepedent measurements of age- and metallicity-sensitive indices. Moreover, as we noted before, the stellar population properties across this relation display a complex behavior that is hard to capture if we analyze projections of it, studying bins of halo mass or stellar mass. 

\subsection{Halo mass estimation}
At the moment, direct halo mass measurements of individual galaxies for large galaxy samples are still scarce. In this sense, one important caveat of our analysis are the halo mass estimates. Halo masses from \citet{2007ApJ...671..153Y} are estimated indirectly using dark matter only simulations, and are based on the luminosity of the groups/clusters. Hence, given that this parameter also correlates strongly with the stellar mass, both axis of the SHMR will depend on the galaxy luminosity. Moreover, for galaxies in isolation or in clusters with few satellites this luminosity is dominated by the luminosity of the central.

Although there are more observational approaches to estimate halo masses for large galaxy samples (see section \ref{intro}), these alternative methods also have associated different uncertainties and systematic errors.  Halo masses can be estimated dynamically in galaxy groups/clusters using their satellite kinematics. However, it is not straight-forward to perform dynamical modelling in halos with only few satellites, given that the velocity dispersion and the extent of the groups in the sky are not clearly defined. Weak gravitational lensing also provides halo mass estimations. Nevertheless, in general, weak lensing measurements refer to galaxy populations rather than individual galaxies. 

Taking all of this into account, we also study the robustness of our results alternative halo mass determinations. We computed the SHMR and the VDHMR using halo masses obtained from different methods (see Appendix~\ref{sec:othercat}). Despite some discrepancies, we generally find the same qualitative trends of the stellar population properties across these relations.

\section{Summary and conclusions}
\label{sec:concl}
We have investigated the stellar population properties of nearby galaxies in terms of their host dark matter halos. Specifically, we derived the average ages, metallicities and [Mg/Fe] abundance ratios. Then, we study these observed galaxy properties across the stellar-to-halo mass relation and the velocity dispersion - halo mass relation, the latter is introduced as a novel approach in this work. Our findings can be summarized as follows:

\begin{itemize}
\item [(i)]
We find smooth trends of age and [M/H] across the SHMR (Fig. \ref{fig:smhm_pops}) and the VDHMR (Fig. \ref{fig:sighm_pops}). At fixed halo mass, we recover classical trends of stellar population studies with galaxies becoming older and more metal-rich with increasing stellar mass or velocity dispersion. Similar behavior is found when using age- and metallicity-sensitive indices, except for [Mg/Fe].

\item [(ii)] 
Besides stellar mass and velocity dispersion, halo mass also drives the stellar populations of galaxies. At low halo masses and at fixed stellar mass or velocity dispersion, galaxies have different stellar population properties depending on the mass of their host halo, with older and more metal-rich galaxies residing in less massive halos than younger and more metal-poor ones. 
\item [(iii)]
Velocity dispersion seems to be the main driver of the stellar population parameters, with halo mass playing a secondary, yet important role determining the stellar population properties of galaxies. We find that velocity dispersion is a better predictor of the stellar population properties than stellar mass. Moreover, we also propose to employ the $\log M_{\star}/R_e-M_h$ relation to compare observations with numerical simulations (assuming that galaxies are virialized systems).

\item [(iv)]
When we probe the VDHMR for early-types and late-types separately, we find very similar trends as the ones shown in Fig. \ref{fig:sighm_pops}, although early-types tend to populate the high-$\sigma_e$ region of the VDHMR with old ages and higher [M/H], while the late-types extend to lower $\sigma_e$ with younger ages and lower [M/H]. The trends observed in halos with low number of galaxies (left panels of Fig. \ref{fig:sighm_pops_env}) also follow the ones seen in Fig. \ref{fig:sighm_pops} (except for the most massive halos). Consistently with the morphology-density relation, the trends for central galaxies in halos with high number of satellites are more similar to the ones seen for early type galaxies (right panels of Fig. \ref{fig:sighm_pops_env}).

\item [(v)]
Our results suggest a connection between the stellar population properties of nearby galaxies and their host dark matter halos. We speculate that the scatter of the ages and metallicities of the galaxies in particular at the low halo mass end is partly driven by the formation times of their host halos, with earlier-formed halos hosting more evolved systems. The trends that we see across the SHMR and VDHMR might be a combined effect of how the assembly of dark matter halos and black hole feedback (probed by the velocity dispersion of the galaxies) affect their host galaxies.
\end{itemize}

Our findings portray a complex scenario in which there are several processes at play regulating the baryonic cycle of galaxies, with a combination of halo growth and black hole feedback driving the evolution of these galaxies. To complement this work, we will study the time evolution of the stellar population properties in a follow-up paper. In order to confirm our proposed scenario, it is important to obtain better halo masses constraints, as well as to understand how other halo properties (i.e., halo formation time, time assembly of dark matter halos) affect the properties of present-day galaxies. Moreover, estimating the halo mass evolution with cosmic time for nearby galaxies, and also studying stellar population properties of higher redshift galaxies in terms of their host halos could also bring light into this matter. Finally, future observations with facilities such as James Webb Space Telescope will be key to understand the connection between black hole growth and galaxy assembly. 


\section*{Acknowledgements}
We are grateful to Carlos Allende for providing the SDSS spectra. We also thank Roberto Maiolino and Jorryt Matthee for their insightful discussions, Marc Huertas-Company for his morphological catalog recommendation and the anonymous referee for her/his  comments, which have contributed to improve this manuscript. We acknowledge support through the RAVET project by the grant PID2019-107427GB-C32 from the Spanish Ministry of Science, Innovation and Universities (MCIU), and through the IAC project TRACES which is partially supported through the state budget and the regional budget of the Consejer\'ia de Econom\'ia, Industria, Comercio y Conocimiento of the Canary Islands Autonomous Community.

We thankfully acknowledge the technical expertise and assistance provided by the 
  Spanish Supercomputing Network (Red Española de Supercomputación) and the Instituto de Astrofísica de Canarias (IAC), as well as the computer 
  resources used: the LaPalma Supercomputer, and the High Performance Computers Diva and Deimos, located at the IAC.

This research made use of Astropy, a community-developed core Python package for Astronomy \citep[][]{astropy:2013,astropy:2018}, and of the Numpy \citep[][]{2020Natur.585..357H}, Scipy \citep[][]{2020SciPy-NMeth} and Matplotlib \citep[][]{2007CSE.....9...90H} libraries.

Funding for SDSS-III has been provided by the Alfred P. Sloan Foundation, the Participating Institutions, the National Science Foundation, and the U.S. Department of Energy Office of Science. The SDSS-III web site is http://www.sdss3.org/.

SDSS-III is managed by the Astrophysical Research Consortium for the Participating Institutions of the SDSS-III Collaboration including the University of Arizona, the Brazilian Participation Group, Brookhaven National Laboratory, Carnegie Mellon University, University of Florida, the French Participation Group, the German Participation Group, Harvard University, the Instituto de Astrofisica de Canarias, the Michigan State/Notre Dame/JINA Participation Group, Johns Hopkins University, Lawrence Berkeley National Laboratory, Max Planck Institute for Astrophysics, Max Planck Institute for Extraterrestrial Physics, New Mexico State University, New York University, Ohio State University, Pennsylvania State University, University of Portsmouth, Princeton University, the Spanish Participation Group, University of Tokyo, University of Utah, Vanderbilt University, University of Virginia, University of Washington, and Yale University.

\section*{Data Availability}

The SDSS group and cluster catalog is available at https://gax.sjtu.edu.cn/data/Group.html. The morphological catalog is available at \href{https://academic.oup.com/mnras/article/476/3/3661/4848300#supplementary-data}{MNRAS}.




\bibliographystyle{mnras}
\bibliography{references} 




\appendix
\section{Indices across the SHMR and VDHMR}
\label{ap:indices}

\begin{figure*}
    \centering
    \includegraphics[width=1.\textwidth]{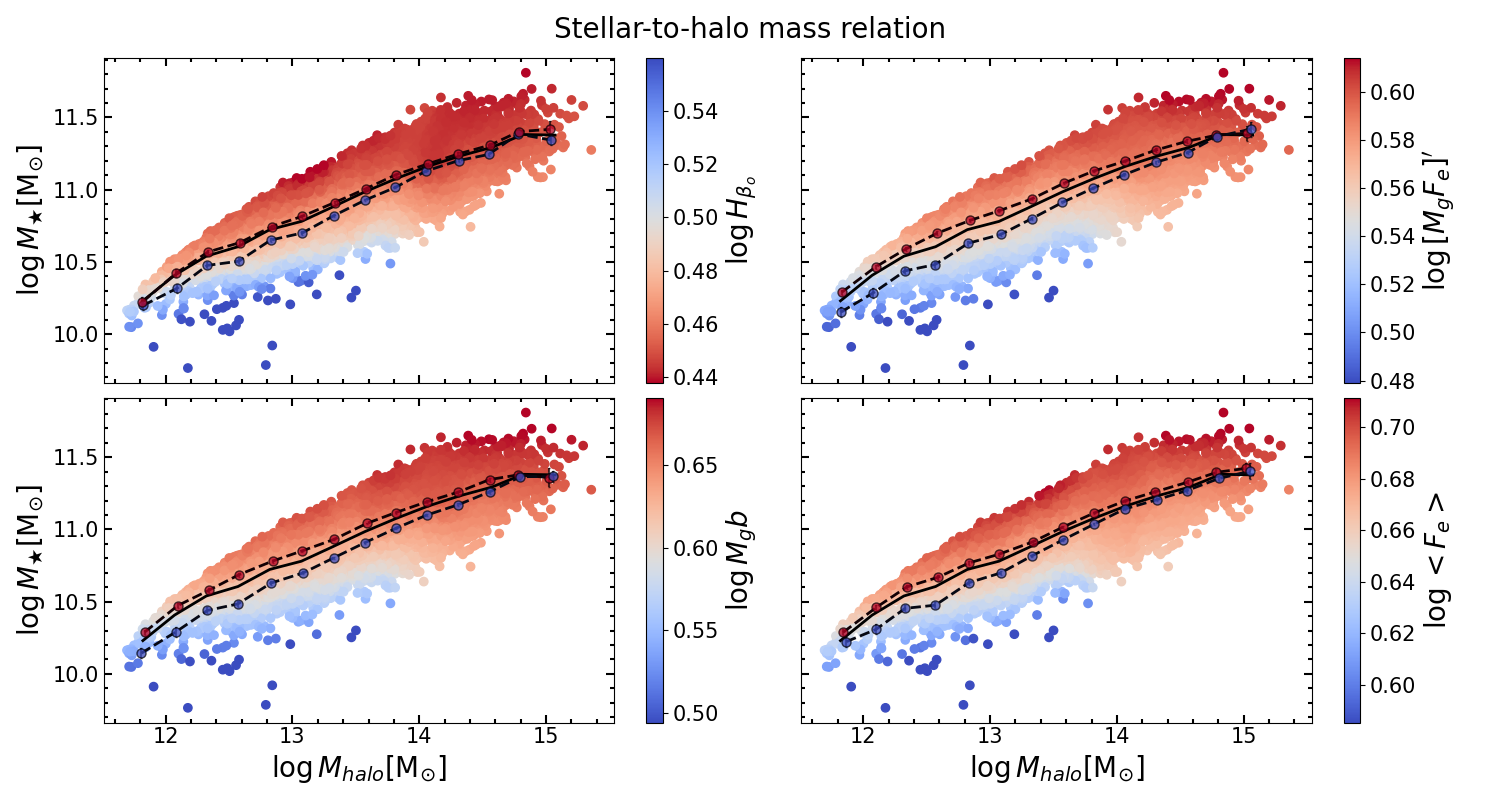}
    \caption{Stellar-to-halo mass relation for SDSS central galaxies color-coded by the age-sensitive index $ \rm H\beta_o$ (upper left), and the metallicity-sensitive indices $\rm [MgFe]^{\prime}$ (upper right), $\rm Mgb5170$ (lower left), $\rm <\!\!Fe\!\!>$ (lower right). We applied the LOESS algorithm to the color-coding in order to highlight the global trends across the SHMR. The median SHMR is shown as a black solid line for reference. At fixed $M_h$, red and blue circles connected by dashed lines indicate the mean $M_{\star}$ of galaxies above the $\rm75^{th}$-percentile and below $\rm25^{th}$-percentile of the corresponding distributions.}
    \label{fig:smhm_indices}
\end{figure*}

\begin{figure*}
    \centering
    \includegraphics[width=1.\textwidth]{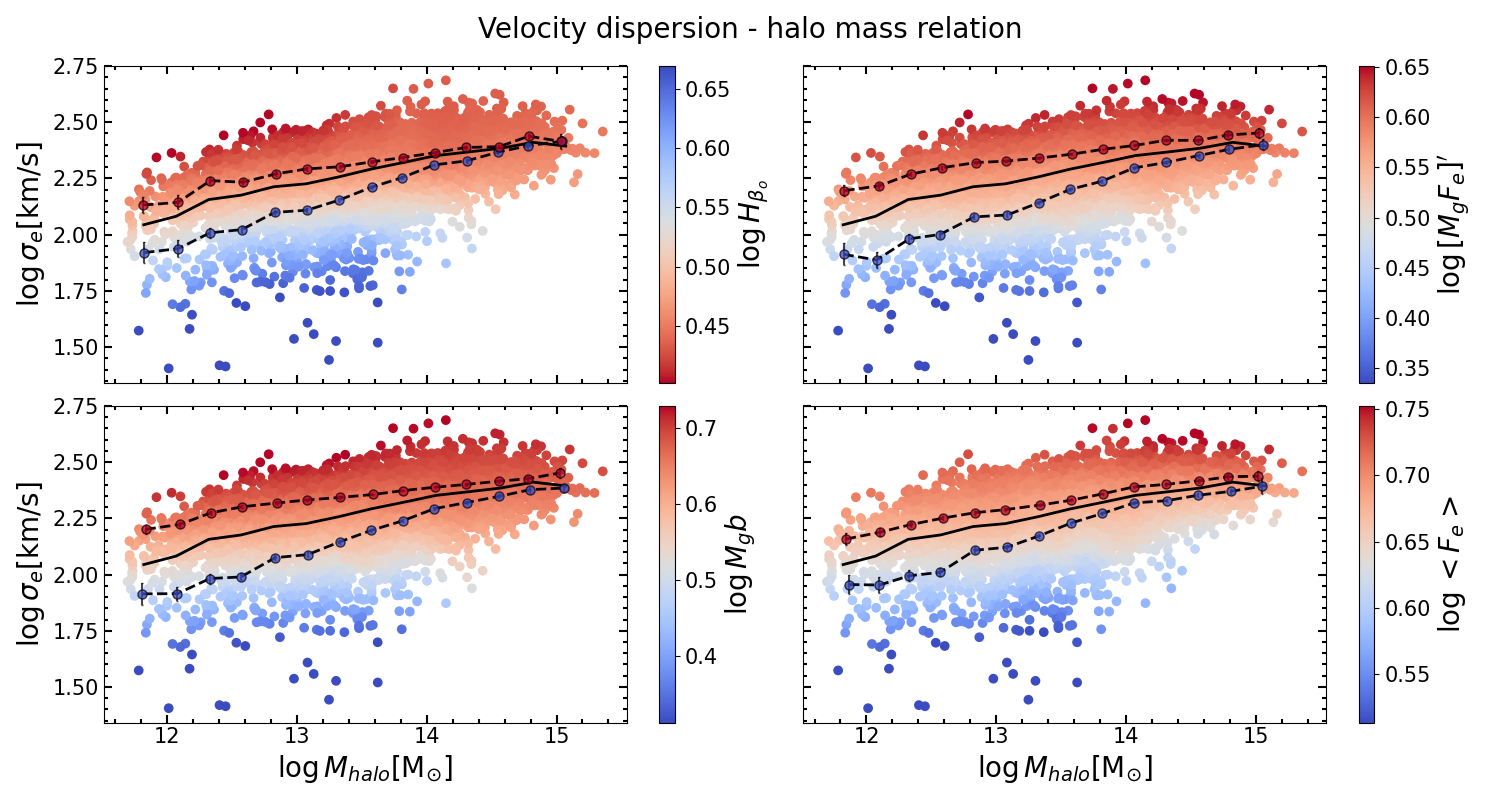}
    \caption{Velocity dispersion (measured within 1 $R_e$) as a function of halo mass for SDSS central galaxies color-coded by the age-sensitive index $ \rm H\beta_o$ (upper left), and the metallicity-sensitive indices $\rm [MgFe]^{\prime}$ (upper right), $\rm Mgb5170$ (lower left), $\rm <\!\!Fe\!\!>$ (lower right).We applied the LOESS algorithm to the color-coding in order to highlight the global trends across the VDHMR. The median VDHMR is shown as a black solid line for reference. At fixed $M_h$, red and blue circles connected by dashed lines indicate the mean $\sigma_e$ of galaxies above the $\rm75^{th}$-percentile and below $\rm25^{th}$-percentile of the corresponding distributions.}
    \label{fig:sighm_indices}
\end{figure*}

In order to test the robustness of our full-spectral fitting approach, we also measured the strengths of absorption features sensitive to age and metallicity in order to have a model-independent first-order estimation of the stellar population parameters. In particular, we use the age-sensitive Balmer index $\mathrm{H\beta_o}$ \citep{2009MNRAS.392..691C}, $\mathrm{Mgb5170}$ \citep{1994ApJS...94..687W}, and the iron indices $\mathrm{Fe5270}$ and $\mathrm{Fe5335}$ \citep{1994ApJS...94..687W}. From that we also calculate the $\mathrm{[MgFe]^{\prime}}$, which is a total metallicity indicator independent of the abundance ratio $\mathrm{[\alpha / Fe]}$ \citep{2003MNRAS.339..897T}. We also compute the combined iron index $\rm <\!\!Fe\!\!>$, as it has smaller errors than the individual iron indices \citep{2000AJ....119.1645T}. In addition, we correct for the LOSVD broadening of the line-strength indices following \citet{2004A&A...426..737K}.

In Fig. \ref{fig:smhm_indices} we show the SHMR color-coded by the age-sensitive index $ \rm H\beta_o$ (upper left), and the metallicity-sensitive indices $\rm [MgFe]^{\prime}$ (upper right), $\rm Mgb5170$ (lower left), $\rm <\!\!Fe\!\!>$ (lower right). We observe how we recover the same trends seen the ages and [M/H] derived with pPXF (see Fig. \ref{fig:smhm_pops}). We also investigate how the indices are displayed across the VDHMR. Figure \ref{fig:sighm_indices} shows the velocity dispersion derived with pPXF (and measured within 1 $R_e$) as a function of halo mass color-coded by the age-sensitive index $ \rm H\beta_o$ (upper left), and the metallicity-sensitive indices $\rm [MgFe]^{\prime}$ (upper right), $\rm Mgb5170$ (lower left), $\rm <\!\!Fe\!\!>$ (lower right). In this case, the trends are also in agreement with the ones derived with pPXF for the age and [M/H] shown in Fig. \ref{fig:sighm_pops}. Taking into account that this approach is model independent and the remarkable agreement between these two methods for both the SHMR and the $\sigma_e-M_h$ relation, we are confident in our results obtained with full-spectral fitting.  Note that in Fig. \ref{fig:smhm_indices} and \ref{fig:sighm_indices} the galaxies for which some of the indices have non-physical values are excluded from the analysis. We also checked that if these galaxies are removed from the main sample, and study the stellar population parameters across the SHMR and the VDHMR, we recover the trends shown in Fig. \ref{fig:smhm_pops} and in \ref{fig:sighm_pops}.

\section{SHMR for late- and early-type galaxies }
\begin{figure*}
    \centering
    \includegraphics[scale = 0.47]{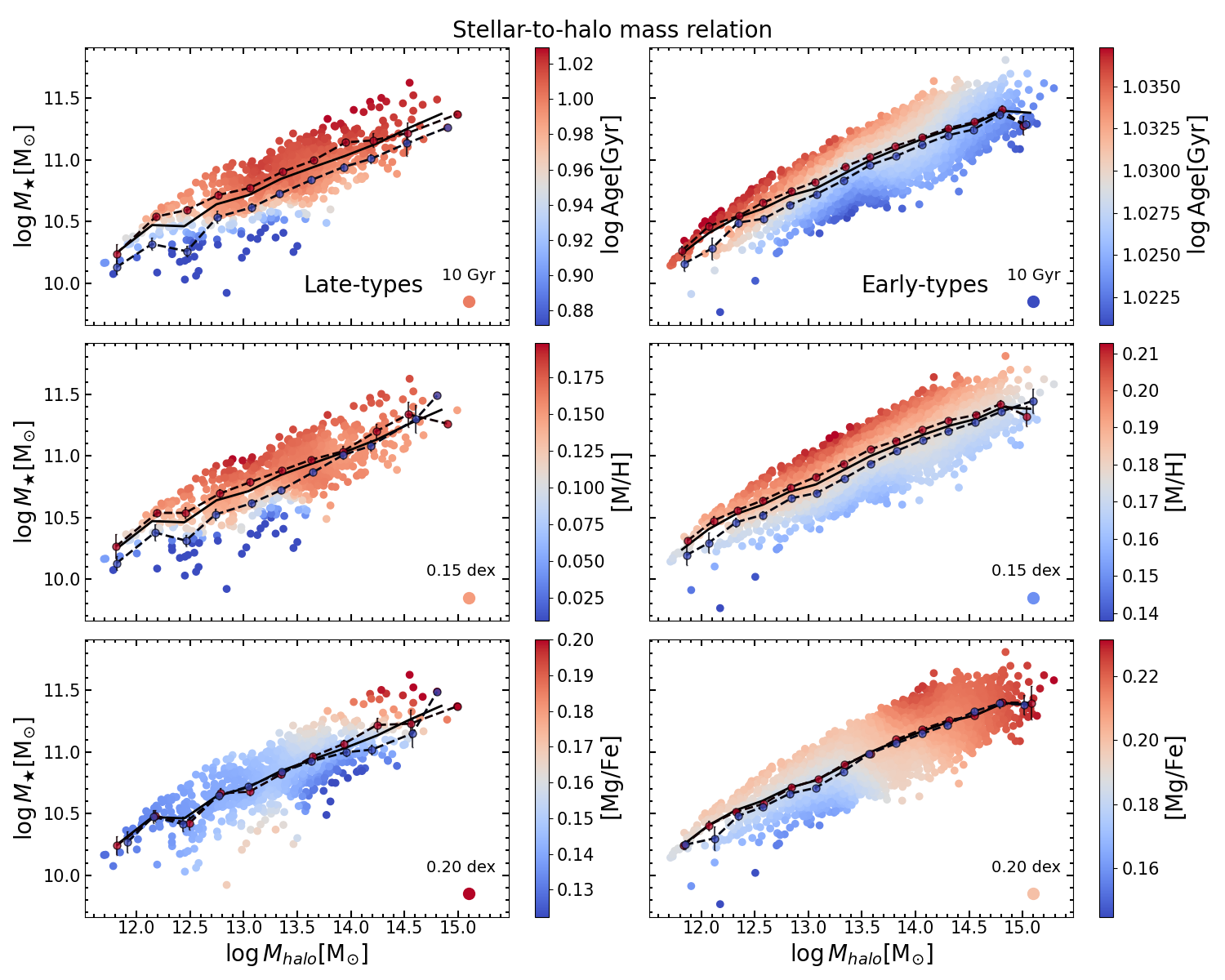}
    \caption{The dependence of the stellar-to-halo mass on morphology. We show the SHMR separately for for late- and early-type galaxies on the left and right columns, respectively. Analogously to figure \ref{fig:smhm_pops}, the SHMR is color-coded by the stellar population properties resulting from our stellar population analysis described in section \ref{an_sp}: mass-weighted ages (upper panels), [M/H] (middle panels) and [Mg/Fe] (lower panels). For reference,  there is a circle in the bottom right corner indicating the color of a galaxy with an age of 10 Gyr (upper panels), metallicity of 0.15 dex (middle panels) and [Mg/Fe] abundance of 0.20 dex. We applied the LOESS algorithm to the color-coding in order to highlight the global trends across the SHMR. The median SHMR is shown as a black solid line for reference. At fixed $M_h$, red and blue circles connected by dashed lines indicate the mean $M_{\star}$ of galaxies above the $\rm75^{th}$-percentile and below $\rm25^{th}$-percentile of the corresponding distributions. At the low halo mass end, late-types extend to younger ages and lower [M/H] than early-types, as the latter tend to populate high stellar mass region with older ages and higher metallicites. We also note the low number of late-types in very massive halos. }
    \label{fig:smhm_pops_morph}
\end{figure*}

We also perform a morphological analysis of the stellar population properties across the SHMR, analogous to the one shown in section \ref{sec:morph} for the VDHMR. In Figure \ref{fig:smhm_pops_morph} we show the age (upper panels), [M/H] (middle panels) and [Mg/Fe] (lower panels) across the VDHMR for late- (left panels) and early-types (right panels), separately.

\label{ap:morph}

\section{Comparison with other halo mass estimates}
\label{sec:othercat}

\begin{figure*}
    \centering
    \includegraphics[scale = 0.25]{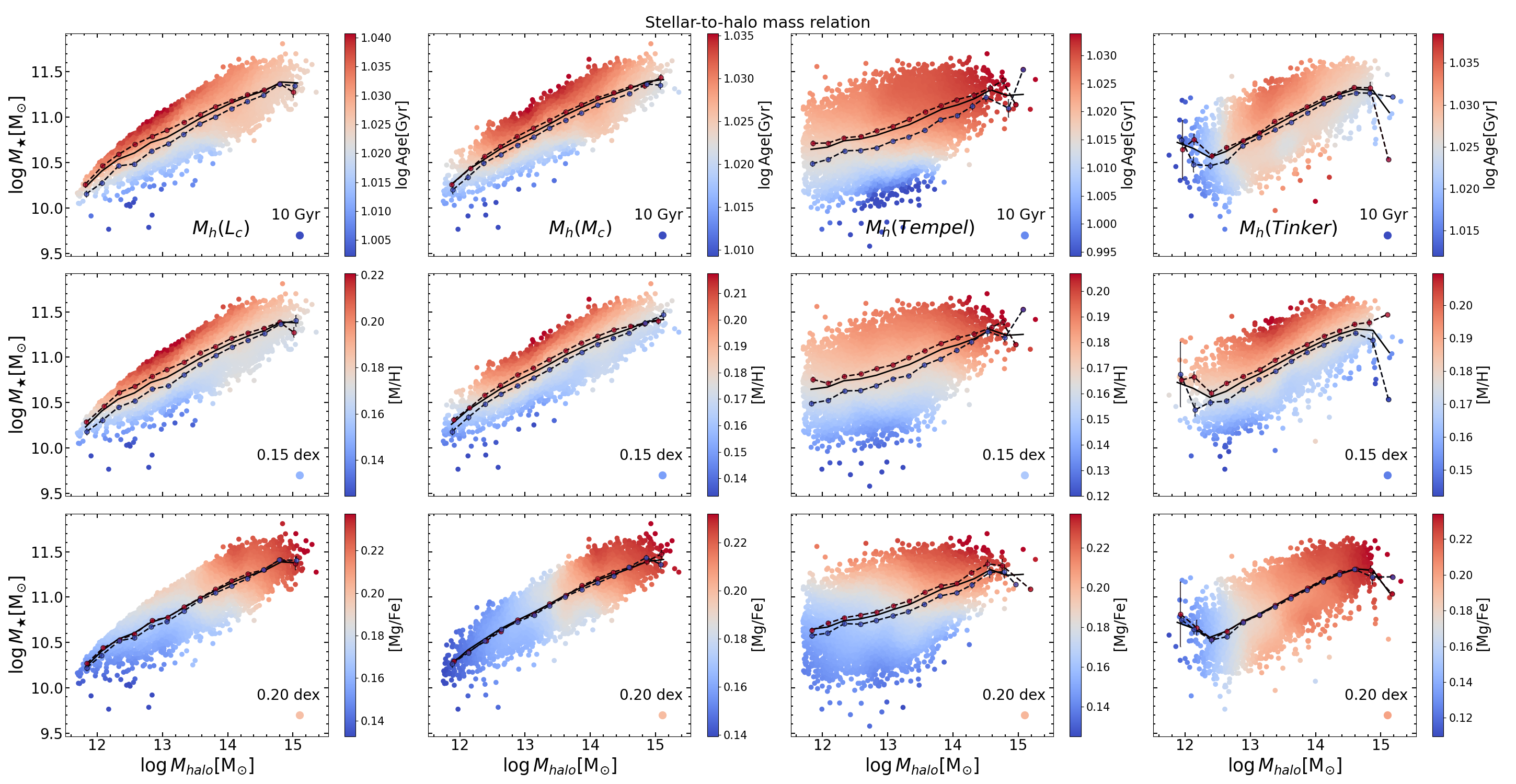}
    \caption{
    Stellar-to-halo mass relation for different halo mass estimates, analogous to Fig. \ref{fig:smhm_pops}, but different columns correspond to different estimates. From left-to-right: $M_h(L_c)$ the halo mass estimate based on $L_c$ from the \citet{2007ApJ...671..153Y} catalog (the same as Fig. \ref{fig:smhm_pops}); $M_h(L_c)$ (the same as Fig. \ref{fig:smhm_pops}); $M_h(M_c)$ (based on $M_c$ from \citet{2007ApJ...671..153Y}); $M_h(Tempel)$, halo mass estimated from \citet{2014A&A...566A...1T}; and $M_h(Tinker)$, halo mass obtained from \citet{2020arXiv201002946T}. Analogously to Fig. \ref{fig:smhm_pops}, the SHMR is color-coded by the stellar population properties resulting from our stellar population analysis described in section \ref{an_sp}: mass-weighted ages (upper panels), [M/H] (middle panels) and [Mg/Fe] (lower panels).  For reference,  there is a circle in the bottom right corner indicating the color of a galaxy with an age of 10 Gyr (upper panels), metallicity of 0.15 dex (middle panels) and [Mg/Fe] abundance of 0.20 dex. We applied the LOESS algorithm to the color-coding in order to highlight the global trends across the SHMR. The median SHMR is shown as a black solid line for reference. At fixed $M_h$, red and blue circles connected by dashed lines indicate the mean $M_{\star}$ of galaxies above the $\rm75^{th}$-percentile and below $\rm25^{th}$-percentile of the corresponding distributions. At the low halo mass end, different halo mass estimates lead to quite different trends of the stellar population parameters.}
    \label{fig:shmr_pops_cat}
\end{figure*}

\begin{figure*}
    \centering
    \includegraphics[scale = 0.25]{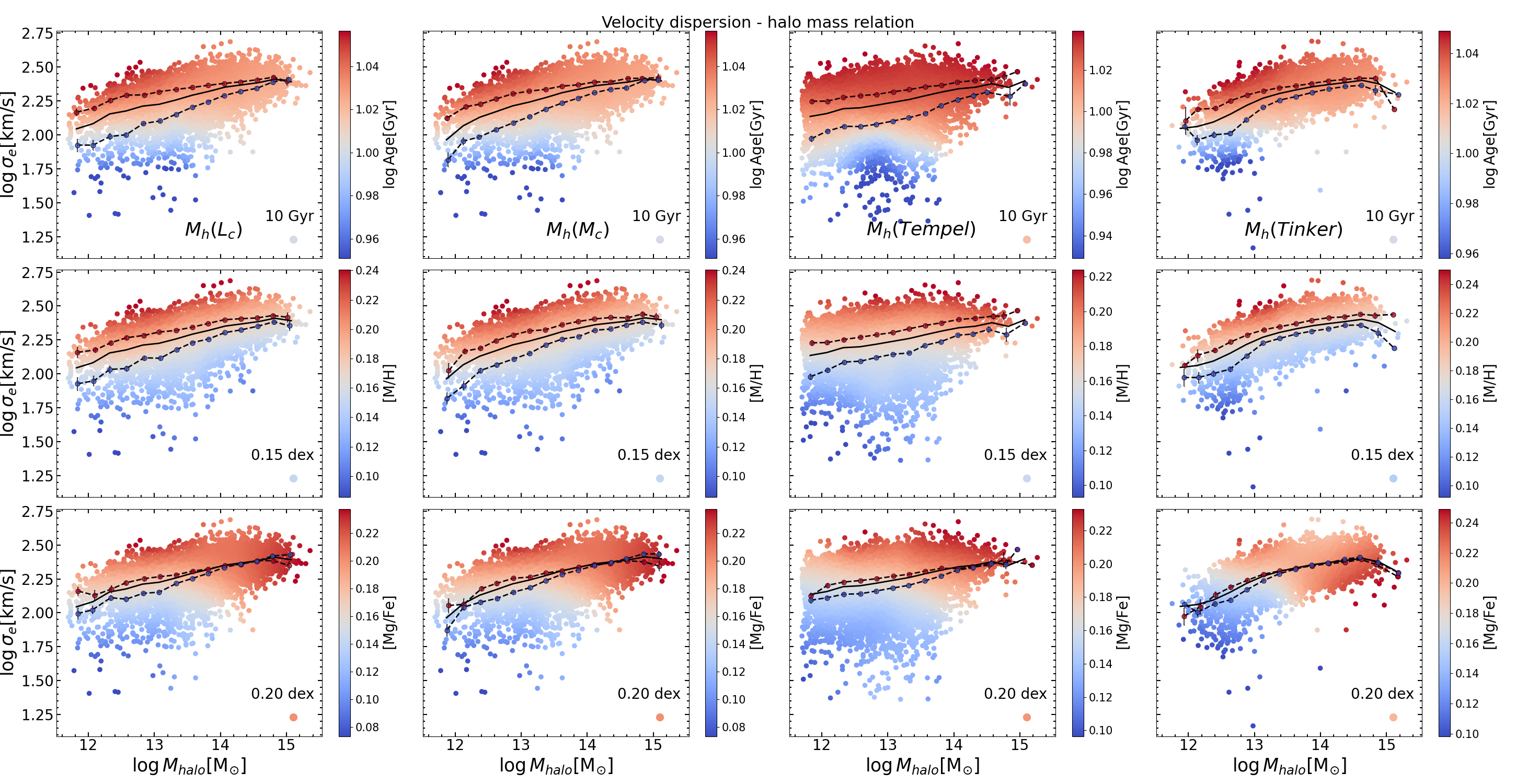}
    \caption{Velocity dispersion - halo mass relation for different halo mass estimates, analogous to Fig.\ref{fig:sighm_pops}, but different columns correspond to different estimates. From left-to-right:  $M_h(L_c)$ the halo mass estimate based on $L_c$  from the \citet{2007ApJ...671..153Y} catalog (the same as Fig. \ref{fig:smhm_pops}); $M_h(L_c)$ (the same as Fig. \ref{fig:smhm_pops}); $M_h(M_c)$ (based on $M_c$ from \citet{2007ApJ...671..153Y}); $M_h(Tempel)$, halo mass estimated from \citet{2014A&A...566A...1T}; and $M_h(Tinker)$, halo mass obtained from \citet{2020arXiv201002946T}. Analogously to Fig. \ref{fig:sighm_pops}, the VDHMR is color-coded by the stellar population properties resulting from our stellar population analysis described in section \ref{an_sp}: mass-weighted ages (upper panels), [M/H] (middle panels) and [Mg/Fe] (lower panels). For reference,  there is a circle in the bottom right corner indicating the color of a galaxy with an age of 10 Gyr (upper panels), metallicity of 0.15 dex (middle panels) and [Mg/Fe] abundance of 0.20 dex. We applied the LOESS algorithm to color-coding in order to highlight the global trends across the VDHMR. The median VDHMR is shown as a black solid line for reference. At fixed $M_h$, red and blue circles connected by dashed lines indicate the mean $\sigma_e$ of galaxies above the $\rm75^{th}$-percentile and below $\rm25^{th}$-percentile of the corresponding distributions. Qualitative trends of the ages and [M/H] of the galaxies across this relation are very similar for different halo mass estimates. }
    \label{fig:sighm_pops_cat}
\end{figure*}

We have analyzed the stellar population properties of 8,778 galaxies across the SHMR and the VDHMR, using halo masses estimated by \cite{2007ApJ...671..153Y} obtained through abundance matching by ranking the characteristic luminosities of the groups (as explained in section \ref{sec:data}). In order to assess the robustness of our results against systematics and uncertainties in the halo mass estimates, in this section we study the SHMR and the VDHMR using halo mass estimates from different sources and/or methodologies.

As described in section \ref{sec:data}, \citet{2007ApJ...671..153Y} estimated halo masses both ranking the total characteristic luminosity, $M_h(L_c)$, and mass of the groups, $M_h(M_c)$. In this section we repeat the analyses shown in sections \ref{sec:smhm} and \ref{sec:sigmh} with $M_h(M_c)$ and halo mass estimates from \citet{2014A&A...566A...1T} and \citet{2020arXiv201002946T} catalogues. We select the galaxies from these catalogues present in our initial sample and impose that $N_{gal}>3$. Our sample has 12,477 galaxies from \citet{2014A&A...566A...1T}, who determines the membership of the galaxies through a modified FoF method, and estimates the masses of the groups dynamically using the virial theorem. We have also 8,234 galaxies in common with \citet{2020arXiv201002946T}, who applies a self-calibrated group finder algorithm, which is an improvement of the \citet{2005MNRAS.356.1293Y} halo-based algorithm. In order to do the comparison, we note that we have selected only galaxies with $M_h>10^{11.7} \, \rm M_{\odot}$.

\subsection{Stellar-to-halo mass relation}
We investigate the SHMR for different halo mass estimates in Fig. \ref{fig:shmr_pops_cat}, where each column is analogous to Fig. \ref{fig:smhm_pops}. From left to right the halo mass estimates correspond to: $M_h(L_c)$ the halo mass estimate based on $L_c$ from the \cite{2007ApJ...671..153Y} catalog (the same as Fig. \ref{fig:smhm_pops}); $M_h(M_c)$ (based on $M_c$ from \cite{2007ApJ...671..153Y}); $M_h(Tempel)$, halo mass estimated from \citet{2014A&A...566A...1T}; and $M_h(Tinker)$, halo mass obtained from \citet{2020arXiv201002946T}. For reference, there is a circle in the bottom right corner indicating the color of a galaxy with an age of 10 Gyr (upper panels), metallicity of 0.15 dex (middle panels) and [Mg/Fe] abundance of 0.20 dex.

In the main analysis we first chose the halo mass estimate that is based on the characteristic luminosity $(L_c)$ of the group rather than the one based on the characteristic mass of the group $(M_c)$. This was motivated by the fact that luminosities are directly observed while stellar masses used by \citet{2007ApJ...671..153Y} are derived using the relation between colors and mass-to-light ratios from \citep{2003ApJS..149..289B}. The comparison of the SHMR for these two estimates is shown in the first two columns of Fig. \ref{fig:shmr_pops_cat}, and we observe how the qualitative trends of the stellar population parameters across this relation are quite similar.

In addition, we note that the shape of SHMR relation is considerably different when using different catalogues (the third and forth columns of Fig. \ref{fig:shmr_pops_cat}) in comparison to the SHMR for the \citet{2007ApJ...671..153Y} halo masses described before. There is also a lack of galaxies at the very massive end in these samples. However, the trends of the stellar population parameters across the relation are qualitatively similar. At low halo masses, note that the trends of age for $M_h(Tinker)$ (forth column) are the ones that differ the most, as all galaxies tend to be young regardless of $M_{\star}$. In addition, the [Mg/Fe] abundance seems mainly driven by halo mass.

\subsection{Velocity dispersion - Halo mass relation}
As we mentioned before, $\sigma$ is a better predictor of the stellar population properties of the galaxies and in section \ref{sec:sigmh} we explored these galaxy properties across the VDHMR. In this section we repeat this analysis using different halo masses estimated differently. Analogously to Fig. \ref{fig:shmr_pops_cat}, Fig. \ref{fig:sighm_pops_cat} shows the VDHMR for different halo mass estimates. The first two columns of  Fig. \ref{fig:sighm_pops_cat} show the halo mass estimates from \citet{2007ApJ...671..153Y}. The stellar population properties across the VDHMR show very similar trends as before when using $M_h(M_c)$ for the whole halo mass range.

We also study the VDHMR with halo masses estimated from other catalogues (third and forth columns of \ref{fig:sighm_pops_cat}).The shape of the VDHMR using the different $M_h$ estimates differ less than in the case of the SHMR.  We observe how the general trends of age and metallicity across the VDHMR in the different catalogues are also quite similar, and show the same qualitative behavior.

\paragraph*{} Although the overall shapes of the SHMR and the VDHMR relation differ when the different halo mass estimates are used, the trends of the stellar population parameters across the relations are generally in agreement qualitatively across the relations despite the considerable different methods to obtain halo masses.

\section{SHMR and VDHMR for the whole galaxy sample}
We also repeated the analysis shown in Figures \ref{fig:smhm_pops} and \ref{fig:sighm_pops} for our initial galaxy sample, where we do not impose the cut on the number of galaxies within the groups/clusters. For a better visualization we selected galaxies with $M_h \leq 10^{15} \, M_{\odot}$ and   $M_{\star} \geq 10^{9.5} \, M_{\odot}$  due to the incompleteness of the sample beyond these limits. We end up with a sample of 271,600 galaxies. 

\subsection{SHMR}
\begin{figure}
    \centering
    \includegraphics[scale = 0.445]{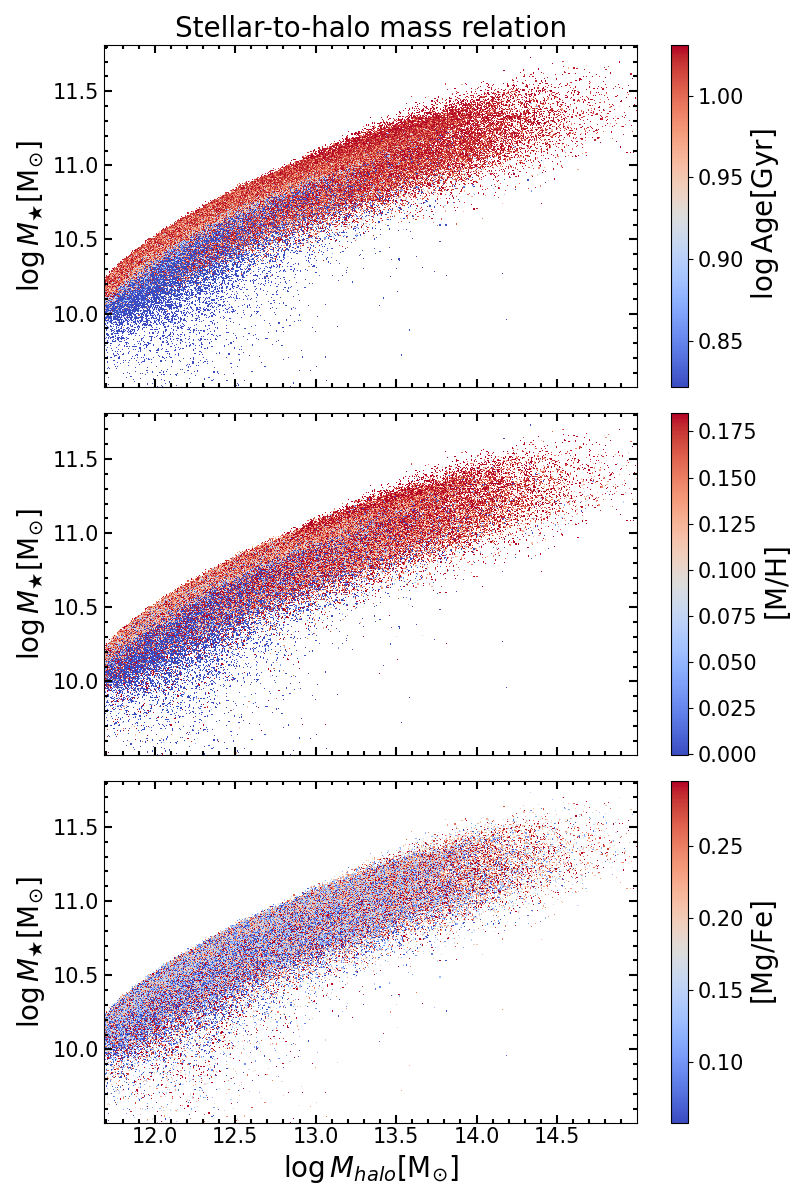}
    \caption{Stellar-to-halo mass relation for SDSS central galaxies color-coded by the stellar population properties resulting from our stellar population analysis described in section \ref{an_sp}: mass-weighted ages (upper panel), metallicities (middle panel) and [Mg/Fe] abundance (bottom panel). The median of these parameters is shown over  stellar mass and halo mass bins of $0.001\times0.001$ dex.}
    \label{fig:smhm_pops_ap}
\end{figure}

In Fig. \ref{fig:smhm_pops_ap}, we show the SHMR color-coded by the mass-weighted ages (upper panel), metallicities (middle panel) and [Mg/Fe] abundance ratios (bottom panel) resulting from our stellar population analysis described in section \ref{an_sp}. The median of these parameters is shown over stellar mass and halo mass $0.001\times0.001$ dex bins, as our sample comprises a large number of galaxies overlapping with similar $M_{\star}$ and $M_h$. In this case, we did not apply the LOESS algorithm given the large number of galaxies of this sample.
For age and [M/H], we observe very similar trends to the ones shown in Fig. \ref{fig:smhm_pops}. The behavior of the [Mg/Fe] abundances is less clear, although we observe slight trend with halo mass similar to the one of \ref{fig:smhm_pops}. 

\subsection{VDHMR}
\begin{figure}
    \centering
    \includegraphics[scale = 0.445]{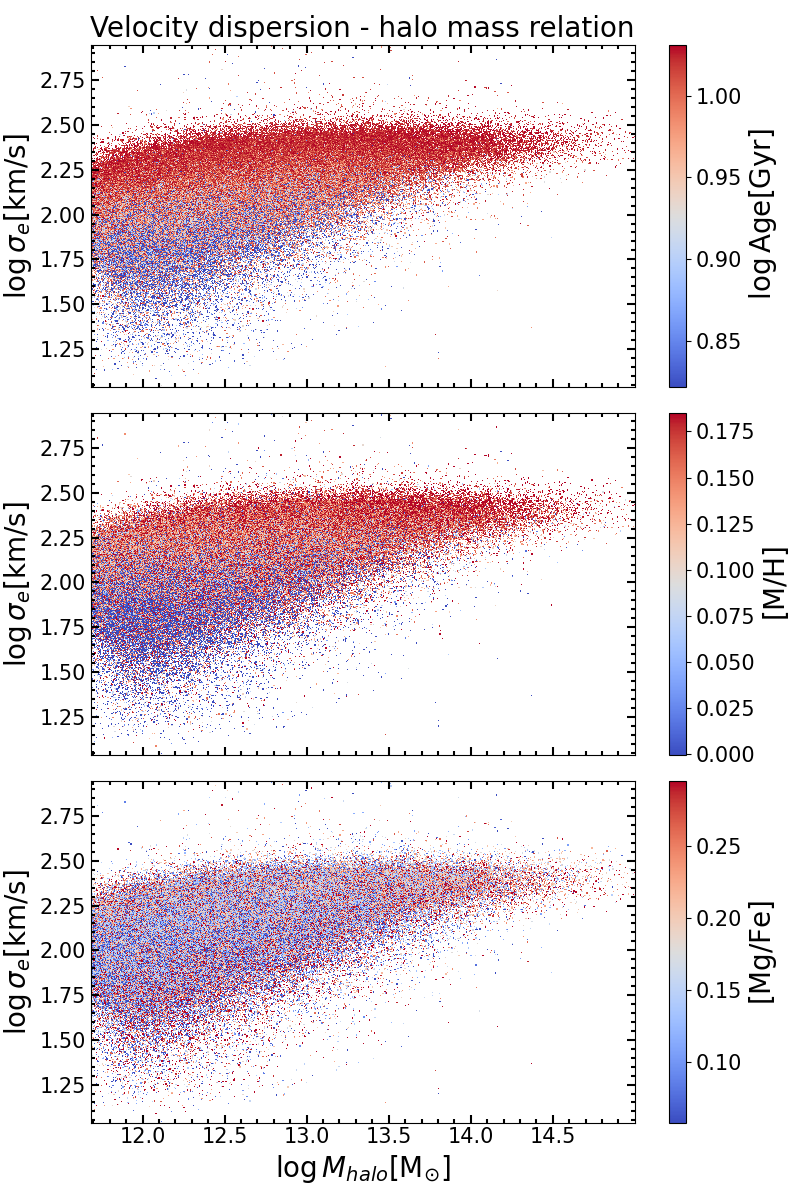}
    \caption{Velocity dispersion (measured within 1 $R_e$) as a function of halo mass for SDSS central galaxies color-coded by the stellar population properties resulting from our stellar population analysis described in section \ref{an_sp}: mass-weighted ages (upper panel), metallicities (middle panel) and [Mg/Fe] abundance (bottom panel). We show the median of these parameters over velocity dispersion and halo mass bins of $0.001\times0.001$ dex.  }
    \label{fig:sighm_pops_ap}
\end{figure}

Fig. \ref{fig:sighm_pops_ap} shows the velocity dispersion (measured within the SDSS fiber and corrected to 1 $R_e$ as detailed in section \ref{an_kine}) as a function of halo mass and color-coded by the mass-weighted ages (upper panel), metallicities (middle panel) and [Mg/Fe] abundance ratios (bottom panel) resulting from our stellar population analysis described in section \ref{an_sp}. The median of these parameters is shown over  velocity dispersion and halo mass $0.001\times0.001$ dex bins.

Similar to what we found for the SHMR, the trends of age and [M/H] are in general agreement with the ones shown in Fig. \ref{fig:sighm_pops}, although there are some differences. We observe different regions that seem to be slightly more Mg-enhanced. We also see that galaxies in the most massive halos also have higher [Mg/Fe] abundances. Moreover, at low halo masses, we also see that galaxies with low-$\sigma_e$ are also slightly more $\alpha$-enhanced. In this regime, there is also a small region where galaxies in the less massive halos and with very high-$\sigma_e$ seem to have higher [Mg/Fe] ratios. However, we note that these trends are not very clear and are considerably mild.

\bsp	
\label{lastpage}
\end{document}